\documentclass[aip,jcp,reprint,noshowkeys,superscriptaddress,twocolumn]{revtex4-1}
\usepackage{graphicx,dcolumn,bm,xcolor,microtype,multirow,amsmath,amssymb,amsfonts,physics,mhchem,xspace,subfigure}

\usepackage[utf8]{inputenc}
\usepackage[T1]{fontenc}
\usepackage{txfonts}

\usepackage[colorlinks=true,citecolor=blue,
    breaklinks=true 
	]{hyperref}
\usepackage[normalem]{ulem}

\newcommand{\matelem}[3]{\langle #1 | #2 | #3 \rangle}
\newcommand{\deriv}[3]{\frac{\partial^{#3} #1}{\partial {#2}^{#3}}}
\newcommand{\bd}[1]{{\bf {#1}}}
\newcommand{\br}[0]{{\bf {r}}}
\newcommand{\bri}[1]{{\bf r}_{#1}}

\newcommand{\frogg}[0]{\text{FROGG}}
\newcommand{\mfrogg}[0]{\mu^\text{FROGG}}
\newcommand{\dr}[1]{\text{d}{\bf r_{#1}}}
\newcommand{\psiex}[0]{\Psi^{\text{ex}}_0}
\newcommand{\phiex}[0]{\Phi^{\text{ex}}_0}
\newcommand{\phimu}[0]{\Phi^{\text{ex}}_0[\mu]}
\newcommand{\phimui}[0]{\Phi^{\text{ex}}_i[\mu]}
\newcommand{\phimub}[0]{\Phi^{\mathcal{B}}_0[\mu]}

\newcommand{\psimub}[0]{\Psi^{\mathcal{B}}_0[\mu]}
\newcommand{\phiimub}[0]{\Phi^{\mathcal{B}}_i[\mu]}
\newcommand{\basis}[0]{\mathcal{B}}

\newcommand{\muueg}{\mu_{\text{UEG}}}
\newcommand{\muuegav}{\langle \mu_{\text{UEG}}\rangle}
\newcommand{\muav}{\langle \mu\rangle}
\newcommand{\mursc}{ \mu_{r_{s,c}}}
\newcommand{\murscav}{\langle \mu_{r_{s,c}}\rangle}
\newcommand{\mursclda}{\langle \mu_{r_{s,c}^{\text{UEG}}}\rangle}

\begin{document}	

\title{A new form of transcorrelated Hamiltonian inspired by range-separated DFT}

\author{Emmanuel Giner}
\email{eginer@lct.jussieu.fr}

\begin{abstract}
The present work introduces a new form of explicitly correlated factor in the context of the transcorrelated methods. 
The new correlation factor is obtained from the $r_{12} \approx 0$ mathematical analysis of the transcorrelated Hamiltonian, and its analytical form is obtained such that the leading order in $1/r_{12}$ of the scalar part of the effective two-electron potential reproduces the long-range interaction of the range-separated density functional theory. 
The resulting correlation factor exactly imposes the cusp and is tuned by a unique parameter $\mu$ which controls both the depth of the coulomb hole and its typical range in $r_{12}$. 
The transcorrelated Hamiltonian obtained with such a new correlation factor has a straightforward analytical expression  depending on the same parameter $\mu$, and its physical contents continuously change by varying $\mu$ : one can change from a non divergent repulsive Hamiltonian at large $\mu$ to a purely attractive one at small $\mu$. 
We investigate the convergence of the ground state eigenvalues and right-eigenvectors of such new transcorrelated Hamiltonian as a function of the basis set and as a function of $\mu$ on a series of two-electron systems. 
We found that the convergence towards the complete basis set is much faster for a quite wide range values of $\mu$. 
We also propose a specific value of $\mu$ which essentially reproduce the results obtained with the frozen Gaussian geminal introduced by Ten-No [CPL-330,169 (2000)]. 

\end{abstract}

\maketitle
\section{Introduction}
One of the most challenging problem in computational chemistry is the accurate simulation of the electronic structure of atomic and molecular systems, which implies solving the Schroedinger equation for a general $N$-body system. 
At the heart of the complexity of such a task relies the rapidly prohibitive computational cost of the wave function methods (WFT), which is induced by the exponential growth with the system size of the Hilbert space involved in the linear eigenvalue problem to be solved. 

In usual WFT, the computational bottleneck is mainly determined by two factors: the level of complexity of the wave function which is imposed by the amount of strong correlation effects in the system, and the convergence of the computed quantities with respect to the size of the one-electron basis used to project the Hamiltonian into a finite eigenvalue problem. 
Typically, strong correlation effects appear when the two-electron coulomb interaction can no longer be considered as a small perturbation over a mean field Hamiltonian. Treating such effects in a black box way involves the use of very flexible wave functions such as 
selected CI\cite{bender,HurMalRan-JCP-73,buenker1,buenker-book,three_class_CIPSI,harrison,cele_cipsi_3_spaces,cele_cipsi_zeroth_order,GinSceCaf-CJC-13, GinSceCaf-JCP-15, ShaHolJeaAlaUmr-JCTC-17}, 
full configuration Quantum Monte Carlo\cite{BooThoAla-JCP-09,BooAla-JCP-10,BooCleThoAla-JCP-11,GhaLozAla-JCP-19,VitAlaKat-JCTC-20} (FCI-QMC) or matrix product states\cite{ChaSha-AR-11,BaiRei-JCP-20}. These methods all have one thing in common: they manage to select, although through different mathematical frameworks, the part of the exponentially growing Hilbert space which dominates the wave function. 

Even if these schemes are extremely efficient to obtain a qualitative description of the wave function through a meaningful selection in an exponentially growing Hilbert space, a quantitative description is often out of reach as it implies the use of very large one-electron basis sets which are needed to recover the slowly converging correlation effects near the electron-electron coalescence point. 
These short-range correlation effects are mainly due to the diverging character of the coulomb potential, which induces a non analytical behaviour of the wave function at small inter-electronic distance ($r_{12}\approx 0$): the famous electronic cusp originally derived by Kato\cite{Kat-CPAM-57}. For a clear and general derivation of the cusp conditions up to second order for Coulombic systems, see Ref. \onlinecite{Tew-JCP-08}. As shown by the seminal works of Hylleraas\cite{Hyl-ZP-29} and further developed by Kutzelnigg and coworkers, \cite{Kut-TCA-85,KutKlo-JCP-91, NogKut-JCP-94} the slow convergence of the correlation energy with respect to the quality of the basis set is mainly due to the impossibility to represent the cusp in a finite one-electron basis set. 
As the cusp conditions originate from the divergence of the Coulomb potential at $r_{12}=0$, an alternative approach would be to develop theories dealing with a smooth potential. Such a path was followed by three distinct branches: 
the explicitly correlated methods (F12), transcorrelated methods (TC) and the range-separated density functional theory (RS-DFT). 

In RS-DFT\cite{Sav-INC-95,TouColSav-PRA-04}, the electron-electron interaction is split into a smooth long-range part and a complementary short-range diverging part, the former being treated explicitly by a wave function and the latter by a density functional. 
Such a splitting is done through the function $\text{erf}(\mu r_{12})$ where $\mu$ is the range separation parameter which allows one to continually move from Kohn-Sham DFT ($\mu=0$) to pure WFT ($\mu=\infty$). 
As the effective Hamiltonian obtained in RS-DFT is smooth at $r_{12}=0$ for a finite value of $\mu$, the corresponding eigenfunction has no electron-electron cusp\cite{GorSav-PRA-06} and therefore the convergence of the results with respect to the basis set are exponential\cite{FraMusLupTou-JCP-15}. 
A number of approximate RS-DFT schemes have been developed involving single-reference\cite{AngGerSavTou-PRA-05, GolWerSto-PCCP-05, TouGerJanSavAng-PRL-09,JanHenScu-JCP-09,TouZhuSavJanAng-JCP-11, MusReiAngTou-JCP-15,KalTou-JCP-18,KalMusTou-JCP-19} and multi-reference\cite{LeiStoWerSav-CPL-97, FroTouJen-JCP-07, FroCimJen-PRA-10, HedKneKieJenRei-JCP-15, HedTouJen-JCP-18, FerGinTou-JCP-18} WFT methods. 
Nevertheless, there are still some open issues in RSDFT, such as remaining fractional-charge and fractional-spin errors in the short-range density functionals \cite{MusTou-MP-17} or the dependence of the quality of the results on the value of the range-separation parameter $\mu$. 

Another approach to tackle the problems of WFT related to the electron cusp have been proposed by the F12 methods which introduce a function explicitly depending on the inter-electronic coordinate\cite{Ten-TCA-12,TenNog-WIREs-12,HatKloKohTew-CR-12, KonBisVal-CR-12, GruHirOhnTen-JCP-17, MaWer-WIREs-18} (called a Jastrow factor) to describe short-range correlation effects which are absent from the finite basis set. Thanks to strong orthogonality, 
all redundant correlation effects between the Jastrow factor and the basis set are projected out of the wave function, 
and therefore the Jastrow factor only deals with correlation effects orthonormal to the basis set. 
The resulting F12 theories, mainly based on perturbation theory and coupled cluster theory, are therefore 
extremely close to their parent WFT theory, with additional contributions involving two- and three-body terms resulting from the use of the Jastrow factor. 
It is important to notice that because of the orthogonalization between the Jastrow and the basis set, 
the complexity of the $N$-body problem within a given basis set is essentially the same between a F12 theory and the parent theory in WFT. Therefore, no strong compression of the electronic wave function is obtained within a given basis set by the F12 theory, 
but a fast convergence toward the complete basis set limit (CBS) is nevertheless obtained for correlation energies and other properties sensitive to the basis set.  

An alternative point of view using a Jastrow factor is the so-called transcorrelated (TC) approach where the full effect of the Jastrow factor is incorporated into the calculation through a similarity transformation of the original Hamiltonian by the Jastrow factor. 
Seminal equations were derived by Hirschfelder\cite{Hirschfelder-JCP-63} who obtained an effective non hermitian operator based on a specific form of Jastrow factor, and later on Boys, Handy and co-workers\cite{BoyHan-PRSLA-69,BoyHanLin-1-PRSLA-69,BoyHanLin-2-PRSLA-69} derived the equations of the TC Hamiltonian for a general exponential form of the Jastrow factor which insures size extensivity. 
The obtained TC Hamiltonian contains certain new features with respect to the usual Hamiltonian: the TC Hamiltonian contains an additional effective two- and three-electron scalar potential together with is a non hermitian two-electron differential operator. Because of its non-hermitian nature, the TC Hamiltonian looses the variational principle, and Boys and Handy managed to derive equations to optimize both orbitals and Jastrow parameters for a single Slater determinant. 
Nevertheless, such a constrained form for the Slater part of the wave function together with the loose of variational principle makes it difficult to find a monotonic convergence of computed energies\cite{Handy-MolPhys-71}. 

Later on, Ten-No\cite{TenNo-CPL-00-a} proposed to significantly change the paradigm of Boys and Handy: instead of using a single Slater determinant and optimizing both the orbitals and the Jastrow factor, one uses a more elaborate many body theory with a frozen universal Jastrow factor.  This strategy was developed using as M{\o}ller-Plesset at second order (MP2) in Refs. \onlinecite{TenNo-CPL-00-a,HinTanTen-JCP-01} and a linearised coupled cluster ansatz in later work\cite{HinTanTen-CPL-02}.  
The Jastrow factor introduced by Ten-No, developed as a linear combination of gaussian functions and referred to as the frozen gaussian geminal (FROGG), only depends on the inter-electronic coordinate and is optimized such that the scalar effective two-electron  potential cancels on average the coulomb potential near $r_{12}=0$. The use of gaussian functions to represent the Jastrow factor enabled Ten-No to develop a numerical scheme to evaluate exactly the three-electron integrals present in TC Hamiltonian\cite{TenNo-CPL-00-b}, which is an alternative to the density-fitting method proposed by the same author in the general context of F12 methods\cite{TenMan-JCP-03}. 
In Ref. \onlinecite{HinTanTen-JCP-01}, Ten-No \textit{et. al.} introduced a biorthogonal approach, 
which allows to treat the non-hermitian TC Hamiltonian in a mathematical form more suited 
to develop approaches based on many-body perturbation theory. 
It is also noteworthy that the eigenfunctions of the TC Hamiltonian are invariant by orbital rotation, 
and therefore in the limit where one gives full flexibility to the wave function, 
the biorthogonal and usual orthogonal approaches give the same results. 
Nevertheless, as approximations and truncations are necessary done in the wave function ansatz for realistic systems, 
the biorthogonal approach might improve the quality of the results in practice. 

In later works, Umezawa \textit{et. al.}\cite{UmeTsu-JCP-03,UmeTsuOhnShiChi-JCP-05} introduced a scheme to couple TC equations with the usual variational Monte Carlo (VMC) scheme in order to optimise the Jastrow factor through a variance miminization. 
An attempt to make TC equations variational have been proposed by Luo\cite{Luo-JCP-10,Luo-JCP-11} based on the empirical experience that in the original single determinant TC approach of Boys and Handy, the orbital optimization seemed to be the source of loss of the variational property. Luo therefore proposed to replace the original TC equations to optimize the orbitals by that of a general Jastrow-Slater ansatz in VMC. 
An alternative to the non hermitian nature of TC theory have been proposed by Yanai \textit{et. al.}\cite{YanShi-JCP-12}  where the Jastrow factor is expressed as an anti-hermitian operator, allowing one to obtain an hermitian effective Hamiltonian after the similarity transformation, which is then efficiently truncated through the use the canonical transformation\cite{NeuYanCha-MolPhys-10}. Applications of such method in the context of quantum computing have been recently published\cite{ValTak-PCCP-20}. 

Further developments of the TC method towards the treatment of solid state systems have been carried by Ochi \textit{et. al.}\cite{OchSodSakTsu-JCP-12,OchTsu-JCTC-14,OchTsu-CPL-15,OchYamAriTsu-JCP-16}, which includes both ground states calculations within a single determinant wave function \cite{OchSodSakTsu-JCP-12,OchYamAriTsu-JCP-16} or at the MP2 level\cite{OchTsu-CPL-15}, together with excited states within configuration interaction\cite{OchTsu-JCTC-14}. 
The TC framework has also been used in the context of DFT to develop new approximations of density functionals\cite{ImaScu-JCP-03,UmeChi-PRA-06,Umezawa-JCP-17}. 

More recently, Alavi \textit{et. al}\cite{CohLuoGutDobTewAla-JCP-19} applied the TC equations with the use of elaborate Jastrow factors which explicitly takes into account the electron-electron together with electron-electron-nucleus coordinates. The Jastrow factor used in such work were obtained from the optimized Jastrow factors for He to Ne of Moskowitz \textit{et. al.}\cite{SchMos-JCP-90} in the context of quantum Monte Carlo. 
In contrast to previous works where a constrained form was given to the wave function (such as coupled cluster for instance\cite{HinTanTen-CPL-02}),  Alavi \textit{et. al.} allow a full flexibility to the Slater part of the wave function to adapt to the elaborated Jastrow factors: they fully solve the $N$-electron problem corresponding to the TC Hamiltonian within a given basis set. 
This is done through the use of the FCI-QMC\cite{BooThoAla-JCP-09,BooAla-JCP-10,BooCleThoAla-JCP-11,GhaLozAla-JCP-19,VitAlaKat-JCTC-20} which is a projective technique allowing one to obtain the ground state of an operator, hermitian or not, through a stochastic sampling of the corresponding Hilbert space. The use of a TC Hamiltonian in the context of FCI-QMC has an important advantage: because the TC Hamiltonian already contains the effect of an elaborate Jastrow factor, the right-eigenvectors of the TC Hamiltonian are more compact and therefore the FCI-QMC procedure converges faster. 
This approach share common points with that of Ten-No in the sense that they both use a fixed Jastrow factor and use a flexible form of wave functions, 
even though the FCIQMC is certainly more flexible than the linearised coupled cluster ansatz of Ten-No used in Ref. \onlinecite{HinTanTen-JCP-01}. 

In the present work, we derive a new form of Jastrow factor such that the leading $1/r_{12}$ terms of the corresponding TC Hamiltonian reproduce the RS-DFT effective Hamiltonian. 
As in RS-DFT, the new TC Hamiltonian and Jastrow factor are tuned by a unique range separation parameter $\mu$. 
The aim of this paper is to establish the analytical form of the TC Hamiltonian and perform a numerical study on the ground state eigenvalues and eigenvectors on a set of two-electron atomic and molecular systems: the helium isoelectronic series from H$^-$ to Ne$^{8+}$ and H$_2$ molecule. 
In the context of the TC theory the interesting features of the present work are that 
i) the analytical form of the TC Hamiltonian is explicitly known and it turns out that all two-electron integrals can be performed analytically and the three-electron integrals can be efficiently computed in mixed numerical-analytical way, ii) the new Jastrow factor and corresponding TC Hamiltonian are tuned by a unique parameter $\mu$ which allows some flexibility, iii) there always exists a regime of $\mu$ which significantly improves the basis set convergence of WFT, iv) we propose schemes to find a value of $\mu$ which automatically adapts to the system and improves basis set convergence. 

The paper is organized as follows. 
In the first section, inspired by the RS-DFT effective Hamiltonian,  
we derive the analytical form of a new Jastrow factor tuned by a unique parameter $\mu$, and discuss the physical content of such a new Jastrow factor. 
Then, we show that the corresponding TC Hamiltonian has a straightforward analytical structure, which is briefly reviewed as a function of the $\mu$ parameter. 
We perform a numerical study of the behaviour of the ground state eigenpair of $\tilde{H}[\mu]$ in the case of the helium atom: 
in the first part we analyze the speed of convergence of the ground state energy with respect to the basis set and the parameter $\mu$, 
then we study the behaviour of the right eigenvectors in real space and compare it with the numerically exact ground state wave function. 
We show that the rapid convergence of the total energies coincides with a rapid convergence of the right eigenvectors in real space, and how to recover a very good approximation of the exact ground state wave function. 
Based on such encouraging results, we study the ground state energy of the helium isoelectronic series from H$^-$ to Ne$^{8+}$ and compare different schemes to obtain 
a value of $\mu$ which automatically adapts to the system and still provides a fast convergence of the total energies. 
Eventually, we study the H$_2$ molecule and compare different regimes of correlation. 
Numerical comparison with the TC Hamiltonian obtained with the FROGG is also performed for all systems studied.  

\section{A new form of Jastrow factor for transcorrelated Hamiltonians}
This section is dedicated to the derivation of a new form of Jastrow factor mimicking at short range the effective Hamiltonian of RS-DFT. 
For the sake of simplicity, we derive the main equations in the case of the helium atom which have an explicit form in terms of the $r_{12}$ coordinate. 
Then, we derive the analytical form of the corresponding new transcorrelated Hamiltonians $\tilde{H}[\mu]$ in the case of a general $N$-electron system, and study its physical content as a  function of the parameter $\mu$. 

\subsection{A new form of Jastrow factor $u(\mu,r_{12})$ tuned by a single parameter $\mu$}
\label{sec:new_j}
\subsubsection{A simple physical picture: the transcorrelated Hamiltonian for the helium atom}
\label{sec:he_htilde}
Let us write the Hamiltonian of the helium atom using the $r_i = |{\bf r}_i|$ and $ r_{12} = |{\bf r}_1 - {\bf r}_2 |$ coordinates\cite{Hylleraas-AdvQChem-64}  
\begin{equation}
 H  = h_c + \frac{1}{r_{12}},
\end{equation}
where 
\begin{equation}
 \begin{aligned}
 \label{def:h_c}
 h_c = &-\frac{1}{2} \sum_{i=1}^2 \bigg(\deriv{}{r_i}{2} + \frac{2}{r_i} \deriv{}{r_i}{} + \frac{2 Z}{r_i}\bigg) \\
     &-\bigg( \deriv{}{r_{12}}{2} + \frac{2}{r_{12}} \deriv{}{r_{12}}{} \bigg) \\
     &-\bigg( \frac{\bd{r_1}}{r_1} \cdot \frac{\bd{r_{12}} }{r_{12}}  \deriv{}{r_1}{} + 
                \frac{\bd{r_2}}{r_2} \cdot \frac{\bd{r_{21}} }{r_{21}}  \deriv{}{r_2}{} \bigg).
 \end{aligned}
\end{equation}
\label{sec:he_j}
Now let us consider the transcorrelated Hamiltonian $\tilde{H}[u]$ obtained by the similarity transformation of the usual Hamiltonian by a Jastrow factor $u(r_{12})$ depending only on $r_{12}$: 
\begin{equation}
 \label{eq:ht_0}
 \begin{aligned}
 \tilde{H}[u]&= e^{-u(r_{12})} H e^{u(r_{12})}.
 \end{aligned}
\end{equation}
With respect to $H$, the only additional terms arising in $\tilde{H}[u]$ are those coming from the action of the differential operator in $r_{12}$,
\begin{equation}
 \mathcal{T}[u] =  -e^{-u(r_{12})}\bigg( \deriv{}{r_{12}}{2} + \frac{2}{r_{12}} \deriv{}{r_{12}}{} \bigg)e^{u(r_{12})}.  
\end{equation}
By defining the following operators 
\begin{equation}
 \label{eq:def_tt}
 \tilde{t}[u] = -2 \deriv{u(r_{12})}{r_{12}}{} \deriv{}{r_{12}}{},
\end{equation}
\begin{equation}
 \label{eq:def_wt}
 \tilde{W}[u] = -\frac{2}{r_{12}} \deriv{u(r_{12})}{r_{12}}{}  , 
\end{equation}
\begin{equation}
 \label{eq:def_wt}
 \tilde{w}[u] = -\deriv{u(r_{12})}{r_{12}}{2} - \bigg( \deriv{u(r_{12})}{r_{12}}{} \bigg)^2, 
\end{equation}
one can write the Similarity transformed Hamiltonian  as
\begin{equation}
 \label{eq:ht_general}
 \tilde{H}[u] = h_c + \frac{1}{r_{12}}  + \tilde{W}[u] + \tilde{w}[u] + \tilde{t}[u].
\end{equation}
Therefore, one can see from Eq. \eqref{eq:ht_general} that the two electrons experience a modified scalar potential given by $\frac{1}{r_{12}}  + \tilde{W}[u] + \tilde{w}[u]$, and an additional differential operator $\tilde{t}[u]$, 
the latter making $\tilde{H}[u]$ non-hermitian. 

\subsubsection{The working equation for $u(\mu,r_{12})$ inspired by RSDFT}
Having established the form of $\tilde{H}[u] $ in Eq. \eqref{eq:ht_general}, one can notice that the leading order terms in $1/r_{12}$ are $\frac{1}{r_{12}}  + \tilde{W}[u]$. We want now to impose the form of $u(r_{12})$ such that it mimics, at leading order in $1/r_{12}$, the long-range effective interaction entering in the effective Hamiltonian of the RS-DFT (see Eq. \eqref{eq:cusp_psi_mu_0} in Appendix \ref{sec:cusp}). 
Mathematically, this condition implies that 
\begin{equation}
 \begin{aligned}
 \label{def_j_00}
 \tilde{W}[u] + \frac{1}{r_{12}}&= \frac{\text{erf}(\mu r_{12})}{r_{12}} \\ 
\Leftrightarrow -\frac{2}{r_{12}} \deriv{u(r_{12},\mu)}{r_{12}}{} + \frac{1}{r_{12}} & = \frac{\text{erf}(\mu r_{12})}{r_{12}}, 
 \end{aligned}
\end{equation}
which is equivalent to 
\begin{equation}
 \label{def_j_0}
 \deriv{u(r_{12},\mu)}{r_{12}}{} = \frac{1 - \text{erf}(\mu r_{12})}{2}.
\end{equation}
The solution to Eq. \eqref{def_j_0} is 
\begin{equation}
 \label{eq:def_j}
 u(r_{12};\mu) = \frac{1}{2}r_{12}\bigg( 1 - \text{erf}(\mu r_{12})  \bigg) - \frac{1}{2\sqrt{\pi}\mu}e^{-(r_{12}\mu)^2},
\end{equation}
which defines the new Jastrow factor $u(r_{12};\mu)$ which depends on a unique parameter $\mu$.  

The main differences of the present approach with the FROGG introduced by Ten-No\cite{TenNo-CPL-00-a} is that instead of optimizing the Jastrow factor as a linear combination of Gaussian functions such that $\frac{1}{r_{12}}  + \tilde{W}[u] + \tilde{w}[u] \approx 0$ when $r_{12}=0$ through a least-square fit with a specific weighting function, 
we impose the analytical structure of $u(r_{12};\mu)$ such that i) the sum of all terms in $1/r_{12}$ (\textit{i.e.} $\tilde{W}[u] + \frac{1}{r_{12}}$) exactly provides the non divergent long-range interaction $\text{erf}(\mu r_{12})/r_{12}$ for all $r_{12}$, 
ii) the new Jastrow factor $u(r_{12};\mu)$ is more flexible than the FROGG as it is tuned by a unique parameter $\mu$ which determines both the depth of the hole and its typical range in $r_{12}$.  

\subsubsection{A few properties of $u(r_{12};\mu)$}
The new Jastrow $u(r_{12};\mu)$ factor is tuned by a unique parameter, $\mu$, which have the unit of the inverse of a distance. 
From Eq. \eqref{eq:def_j} we can notice that 
\begin{equation}
 \lim_{r_{12} \rightarrow \infty}u(r_{12};\mu) = 0,
\end{equation}
which means that 
\begin{equation}
 \lim_{r_{12} \rightarrow \infty}e^{u(r_{12};\mu)} = 1,
\end{equation}
and therefore $u(r_{12};\mu)$ impacts only the small $r_{12}$ behaviour,  
which is expected since the equation determining $u(r_{12},\mu)$ is obtained from the small $r_{12}$ limit of the Schroedinger equation (see Eq. \eqref{def_j_00}). 

To better understand the physical content of at $r_{12}\approx 0$,  
let us Taylor expand the function $u(r_{12};\mu)$ 
\begin{equation}
 \label{eq:j_dl}
 u(r_{12};\mu) = -\frac{1}{2\sqrt{\pi}\mu} + \frac{1}{2}r_{12} - \frac{\mu}{2\sqrt{\pi}} r_{12}^2 + O(r_{12}^4).
\end{equation}
A first remark is that $u(r_{12};\mu)$ exactly restores the cusp as
\begin{equation}
 \label{eq:cusp_phi_0}
 \lim_{r_{12}\rightarrow 0} \deriv{}{r_{12}}{}u(r_{12};\mu) = \frac{1}{2},
\end{equation}
implying that the right-eigenvectors of the TC Hamiltonian obtained with $u(r_{12};\mu)$ are necessary cusp less (see Appendix \ref{sec:cusp}). 
The Taylor expansion of Eq. \eqref{eq:j_dl} inserted in the exponential correlating Jastrow factor reads 
\begin{equation}
 \label{eq:bigJ_dl}
 e^{u(r_{12},\mu)}\approx e^{-\frac{1}{2\sqrt{\pi}\mu}} e^{ \frac{1}{2}r_{12}} e^{- \frac{\mu}{2\sqrt{\pi}} r_{12}^2}. 
\end{equation}
From Eq. \eqref{eq:bigJ_dl} one can analyze each term: i) as $e^{-\frac{1}{2\sqrt{\pi}\mu}}<1$, the Jastrow factor digs the Coulomb hole, 
ii) the term in $e^{ \frac{1}{2}r_{12}} $ exactly imposes the cusp; iii) and the last term $e^{- \frac{\mu}{2\sqrt{\pi}} r_{12}^2}$ reshapes the curvature of the Coulomb hole. 

\begin{figure*}
        \includegraphics[width=0.45\linewidth]{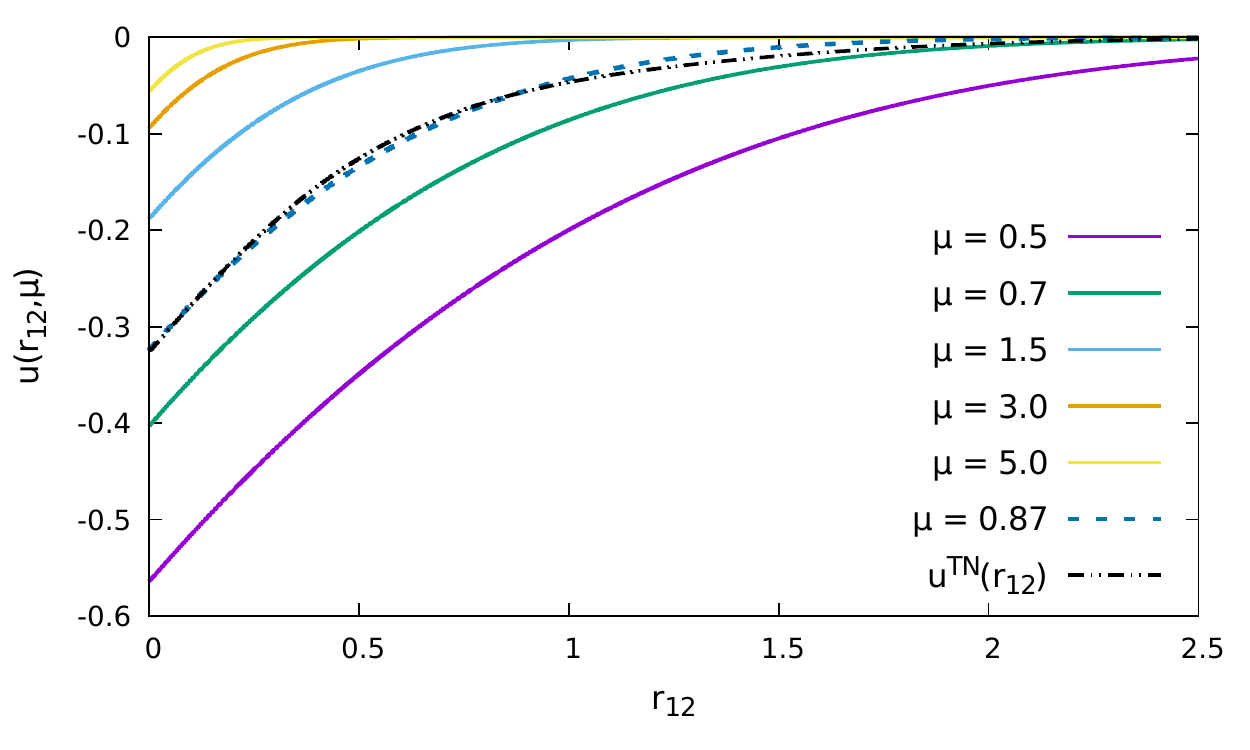}
        \includegraphics[width=0.45\linewidth]{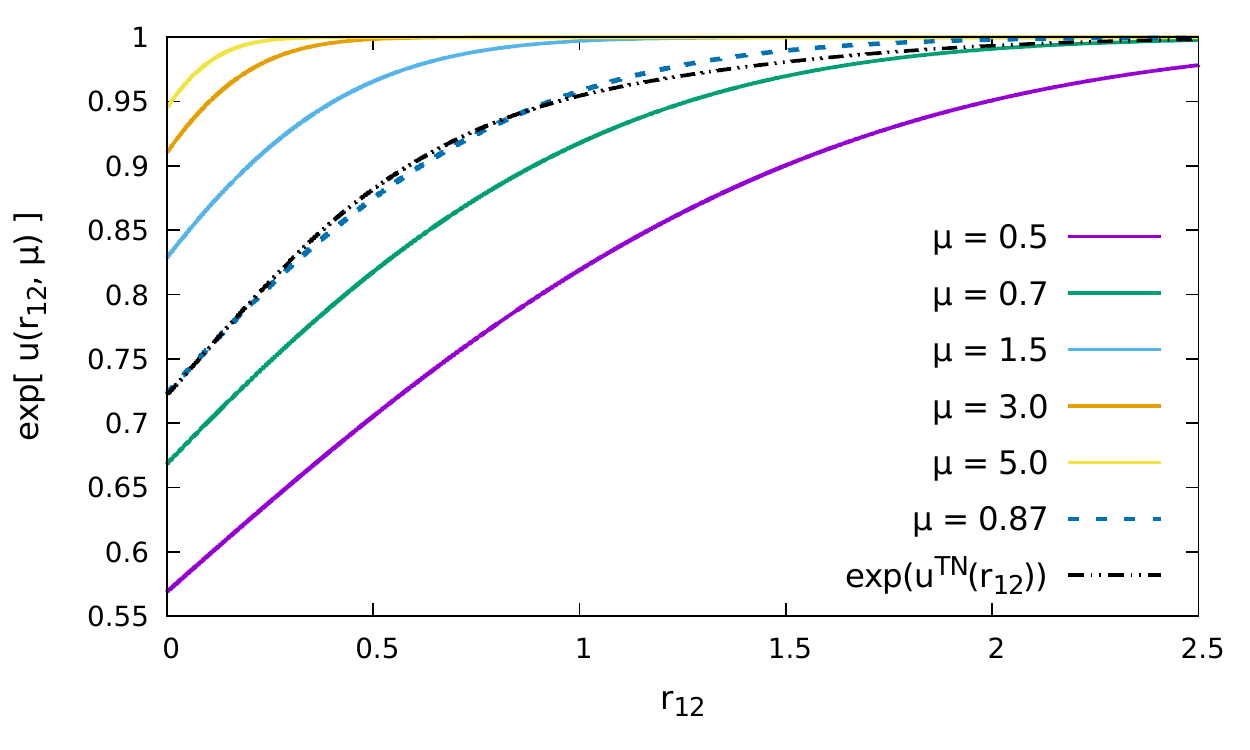}\\
        \caption{Shape of $u(r_{12};\mu)$ (left) and of $\text{exp}\big(u(r_{12};\mu) \big) $ (right) for different values of $\mu$. 
We also compare the FROGG (see Eq. \eqref{eq_frogg}) and the value of $\mu = 0.865$ which equals the FROGG at $r_{12}=0$ (see Eq. \eqref{mu_ten_no_1}). }
 \label{fig_j_mu}
\end{figure*}

Coming now to the behaviour of $u(r_{12};\mu)$ as a function of the parameter $\mu$, we report in Figure \ref{fig_j_mu} its dependence in $r_{12}$ for a set of values of $\mu$, and we compare it with the FROGG. 
From Figure\ref{fig_j_mu}, one can see that: i) the larger the $\mu$, the more short-range is $u(r_{12};\mu)$, ii) the larger the $\mu$, the closer $u(r_{12};\mu)$ is from 0 and therefore the less impact the correlating factor has on the wave function, iii) the smaller the $\mu$, the deeper and broader is the hole imposed by $e^{u(r_{12},\mu)}$. 

As the function $u(r_{12};\mu)$ contains a unique parameter $\mu$, one can try to find the value $\mfrogg$ such that 
$u(r_{12};\mfrogg)$ qualitatively reproduces the FROGG. 
To do so, we impose that $\mfrogg$ is such that the two Jastrow factors coincide at coalescence, \textit{i.e.}  
\begin{equation}
 \label{mu_ten_no_1}
u(r_{12}=0;\mfrogg) = u^{\text{TN}}(r_{12}=0), 
\end{equation}
where $ u^{\text{TN}}(x)$ is the FROGG introduced by Ten-No
\begin{equation}
 \label{eq_frogg}
 u^{\text{TN}}(x) = -\sum_{i=1}^6 c_i e^{-\alpha_i x^2 },
\end{equation}
with the coefficients $c_i$ and exponents $\alpha_i$ are given in Table 1 of Ref. \onlinecite{TenNo-CPL-00-a}. 
Eq. \eqref{mu_ten_no_1} gives 
\begin{equation}
 \label{mu_ten_no_2}
 \mfrogg = -\frac{1}{2\sqrt{\pi}u^{\text{TN}}(0)},
\end{equation}
which gives $\mfrogg\approx0.87$. 
We represent on Fig. \ref{fig_j_mu} the comparison between $u(r_{12};\mfrogg)$ and the FROGG, and it is clearly apparent  that $u(r_{12};\mfrogg)$ reproduces almost perfectly the FROGG, although it decays slightly faster at large $r_{12}$. Therefore, one can see the FROGG as a special case of the new Jastrow factor $u(r_{12};\mu)$ with $\mu = \mfrogg$. 

\subsection{The general analytical form of  $\tilde{H}[\mu]$ and some of its properties}
\label{sec:ht_general}
\subsubsection{General analytical form of $\tilde{H}[\mu]$ for a $N$-electron system}

Having established the analytical expression of $u(r_{12};\mu)$, one can derive the expression of the associated transcorrelated Hamiltonian for a general $N$-electron system.  
Using the general form of the transcorrelated Hamiltonian of Eq. (2) in Ref. \onlinecite{CohLuoGutDobTewAla-JCP-19} and using $u(r_{12},\mu)$ as Jastrow factor gives 
\begin{equation}
 \begin{aligned}
 \label{ht_def_g}
 \tilde{H}[\mu] &\equiv e^{-\hat{\tau}_\mu} \hat{H} e^{\hat{\tau}_\mu} \\ 
                &= H + \big[ H,\hat{\tau}_\mu \big] + \frac{1}{2}\bigg[ \big[H,\hat{\tau}_\mu\big],\hat{\tau}_\mu\bigg],
 \end{aligned}
\end{equation}
with $\hat{\tau}_\mu = \sum_{i<j}u(r_{ij},\mu)$ and $\hat{H} = \sum_i -\frac{1}{2} \nabla^2_i + v(\br{}_i) + \sum_{i<j} \frac{1}{r_{ij}}$. 
Therefore, the \textit{i-th} right-eigenvectors $\ket{\phimui}$ depends on $\mu$ and satisfy the effective Schroedinger equation
\begin{equation}
 \tilde{H}[\mu] \ket{ \phimui}  = E_i \ket{\phimui}, 
\end{equation}
where $E_i$ is the exact $i$-th eigenvalue of the physical Hamiltonian. The fact that $\tilde{H}[\mu]$ and $H$ share the same spectrum originates from the properties of the similarity transformation. Nevertheless, once projected in an incomplete basis set, $\tilde{H}[\mu]$ and $H$ might not have the same eigenvalues. 

Eq. \eqref{ht_def_g} leads to the following transcorrelated Hamiltonian 
\begin{equation}
 \begin{aligned}
 \label{ht_def_gu}
 \tilde{H}[\mu] & = H - \sum_{i<j} \hat{K}(\bri{i},\bri{j},\mu) - \sum_{i<j<k} \hat{L}(\bri{i},\bri{j},\bri{k},\mu),
 \end{aligned}
\end{equation}
where the effective two- and three-body operators $\hat{K}(\bri{1},\bri{2},\mu)$ and $\hat{L}(\bri{1},\bri{2},\bri{3},\mu)$ are defined as
\begin{equation}
 \begin{aligned}
 \label{def_k}
  \hat{K}(\bri{1},\bri{2},\mu)  = \frac{1}{2} \bigg( &\Delta_1 u(r_{12},\mu) + \Delta_2 u(r_{12},\mu) \\
                                               + &\big(\nabla_1 u(r_{12},\mu) \big) ^2 + \big(\nabla_1 u(r_{12},\mu) \big) ^2 \bigg) \\
                                               + &\nabla_1 u(r_{12},\mu) \cdot \nabla_2 + \nabla_2 u(r_{12},\mu)\cdot \nabla_1,
 \end{aligned}
\end{equation}
\begin{equation}
 \begin{aligned}
 \label{def_k}
  \hat{L}(\bri{1},\bri{2},\bri{3},\mu)  = &   \nabla_1 u(r_{12},\mu) \cdot \nabla_1 u(r_{13},\mu) \\
                                          + & \nabla_2 u(r_{21},\mu) \cdot \nabla_2 u(r_{23},\mu)  \\
                                          + & \nabla_3 u(r_{13},\mu) \cdot \nabla_3 u(r_{32},\mu).
 \end{aligned}
\end{equation}

As shown in the Appendix \ref{k_l_appendix}, the operator $\hat{K}(\bri{1},\bri{2},\mu)$ has a straightforward analytical expression which reads 
\begin{equation}
 \label{eq:k_final}
  \begin{aligned}
   \hat{K}(\bri{1},\bri{2},\mu) = & \frac{1 - \text{erf}(\mu r_{12})}{r_{12}} - \frac{\mu}{\sqrt{\pi}} e^{-\big(\mu r_{12} \big)^2} + \frac{\bigg(1 -     \text{erf}(\mu r_{12}) \bigg)^2}{4} \\
& - \bigg( \text{erf}(\mu r_{12}) - 1\bigg) \deriv{}{r_{12}}{},
  \end{aligned}
\end{equation}
and similarly the analytical expression of $\hat{L}(\bri{1},\bri{2},\bri{3},\mu) $ reads 
\begin{equation}
 \label{eq:l_final}
 \begin{aligned}
 \hat{L}(\bri{1},\bri{2},\bri{3},\mu) = & \frac{1 - \text{erf}(\mu r_{12})}{2 r_{12}} \bri{12} \cdot \frac{1 - \text{erf}(\mu r_{13})}{2 r_{13}} \bri{13} \\
                                      + & \frac{1 - \text{erf}(\mu r_{12})}{2 r_{12}} \bri{21} \cdot \frac{1 - \text{erf}(\mu r_{23})}{2 r_{23}} \bri{23} \\
                                      + & \frac{1 - \text{erf}(\mu r_{13})}{2 r_{13}} \bri{31} \cdot \frac{1 - \text{erf}(\mu r_{32})}{2 r_{32}} \bri{32}.
 \end{aligned}                    
\end{equation}

By injecting the form of Eqs. \eqref{eq:k_final} and \eqref{eq:l_final} in Eq. \eqref{ht_def_gu}, one obtains 
\begin{equation}
  \begin{aligned}
 \label{eq:exp_ht_mu}
   \tilde{H}[\mu] = & h_c + \sum_{i<j} \bigg( \tilde{\mathcal{W}}_{ee}(r_{ij},\mu) + \tilde{t}[\mu] \bigg) - \sum_{i<j<k} \hat{L}(\bri{1},\bri{2},\bri{3},\mu)
  \end{aligned}
\end{equation}
where $ \tilde{\mathcal{W}}_{ee}(r_{ij},\mu)$ is the scalar effective two-body potential of  $\tilde{H}[\mu]$
\begin{equation}
 \begin{aligned}
 \tilde{\mathcal{W}}_{ee}(r_{ij},\mu)  =  \frac{\text{erf}(\mu r_{ij})}{r_{ij}} + \frac{\mu}{\sqrt{\pi}} e^{-\big(\mu r_{ij} \big)^2} - \frac{\bigg(1 -     \text{erf}(\mu r_{ij}) \bigg)^2}{4}, 
 \end{aligned}
\end{equation}
and $\tilde{t}[\mu]$ is a non hermitian two-body differential operator 
\begin{equation}
 \begin{aligned}
 \tilde{t}[\mu] =  \bigg( \text{erf}(\mu r_{ij}) - 1\bigg) \deriv{}{r_{ij}}{}.
 \end{aligned}
\end{equation}

\begin{figure*}
        \includegraphics[width=0.45\linewidth]{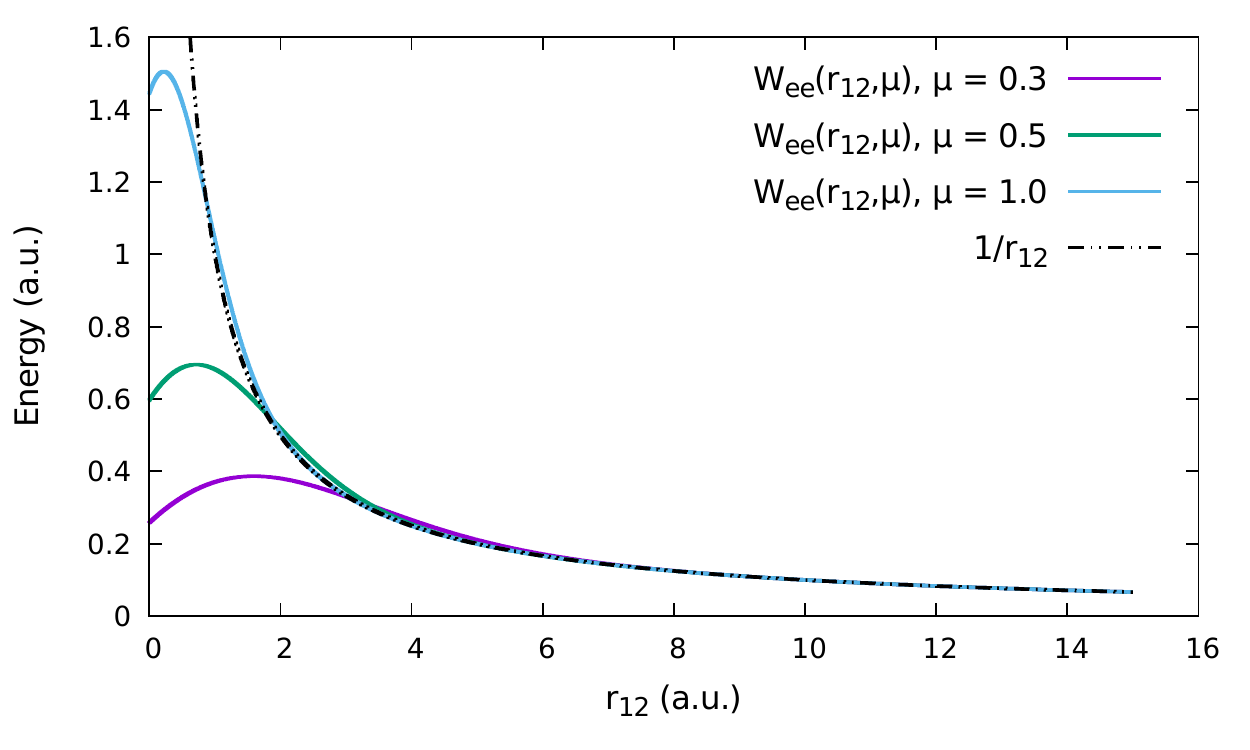}
        \includegraphics[width=0.45\linewidth]{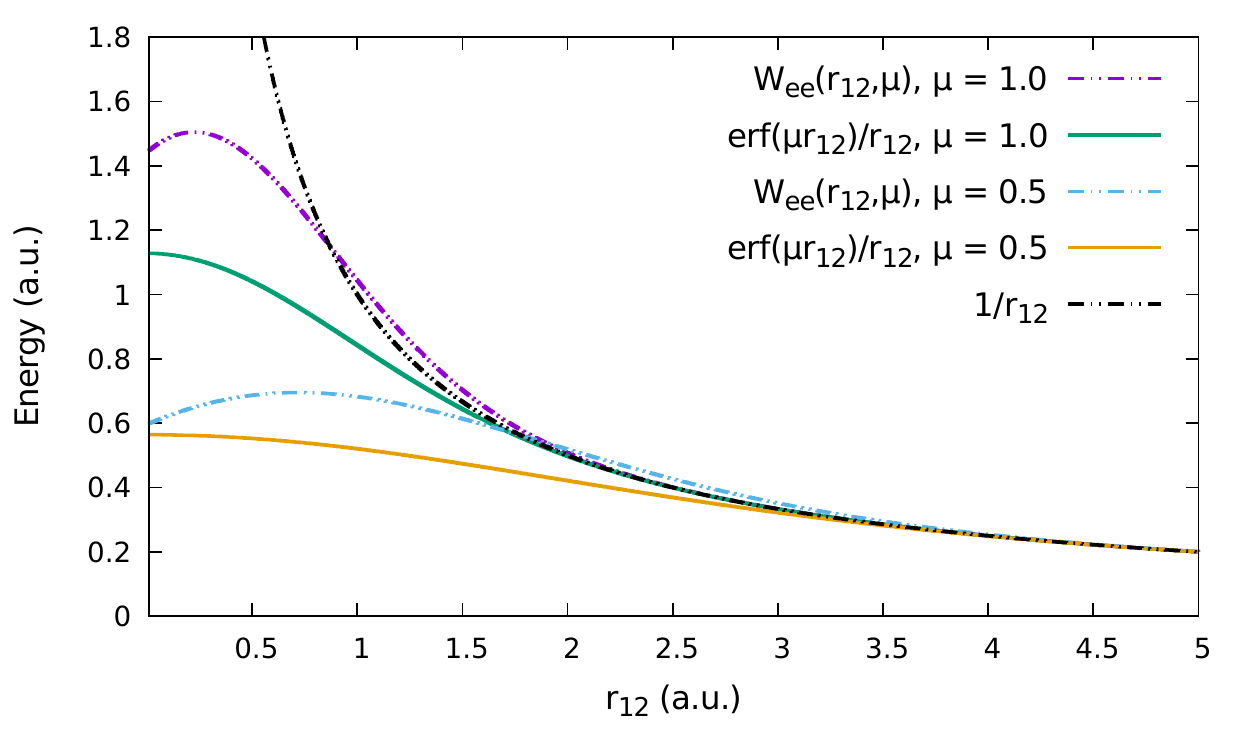}\\
        \caption{Shape of $\tilde{\mathcal{W}}_{ee}(r_{12},\mu)$ (left) and comparison with $\frac{\text{erf}(\mu r_{12}}{r_{12}}$ (right) for different values of $\mu$.}
 \label{fig_wee_mu}
\end{figure*}

As the present study is devoted to two-electron systems, the physics of $\tilde{H}[\mu]$ can be qualitatively understood by looking at $\tilde{\mathcal{W}}_{ee}(r_{ij},\mu)$, which is illustrated in Fig. \ref{fig_wee_mu} for different values of $\mu$ and compared with the usual long-range interaction of the RS-DFT framework. 
From  Fig. \ref{fig_wee_mu} we can see that the scalar effective potential $ \tilde{\mathcal{W}}_{ee}(r_{ij},\mu)$ 
is non divergent, increases when $\mu$ increases and globally resembles the long-range interaction used in RS-DFT at large $r_{12}$. 
Nevertheless, as observed from Fig. \ref{fig_wee_mu}, it is significantly different 
from the long-range interaction at small $r_{12}$, 
and it is not monotonic and attractive (\textit{i.e.} $\lim_{r_{12} \rightarrow 0} \deriv{}{r_{12}}{} \tilde{\mathcal{W}}_{ee}(r_{12},\mu) >0 $). 
This difference with respect to the long-range interaction can be understood as the equation obtained to derive the Jastrow factor only takes into account the first-order derivative of $u(r_{12},\mu)$ (see Eqs. \eqref{def_j_00} and \eqref{def_j_0}). 
If one desires to impose that $\tilde{\mathcal{W}}_{ee}(r_{12},\mu) = \frac{\text{erf}(\mu r_{12})}{r_{12}}$, one would have to solve the non-linear differential equation 
\begin{equation}
 2 \deriv{u(r_{12})}{r_{12}}{} + r_{12} \bigg( \deriv{u(r_{12})}{r_{12}}{2} + \bigg[ \deriv{u(r_{12})}{r_{12}}{} \bigg]^2\bigg) = 1 - \text{erf}(\mu r_{12}), 
\end{equation}
whose solution is unknown to the best of the author knowledge. 

\subsubsection{Limit cases for $\tilde{H}[\mu]$ with $\mu$}
\label{sec:h_mu_lim}
Regarding now the variation of $\tilde{H}[\mu]$, as 
\begin{equation}
 \label{eq:lim_mu_0}
 \lim_{\mu  \rightarrow \infty }u(\mu,r_{12}) = 0, 
\end{equation}
one recovers the usual Hamiltonian in the ${\mu  \rightarrow \infty }$ limit:
\begin{equation}
 \label{eq:lim_mu_1}
 \lim_{\mu \rightarrow \infty} \tilde{H}[\mu] = H,
\end{equation}
although it is only in the sense of distributions (see Appendix \ref{sec:large_mu_lim} for a more detailed explanation). 

In the $\mu \rightarrow 0$ limit, the transcorrelated Hamiltonian $\tilde{H}[\mu]$ becomes simply 
\begin{equation}
 \label{eq:htilde_mu_low}
 \begin{aligned}
&\lim_{\mu \rightarrow 0} \tilde{H}[\mu] = h_c - \frac{1}{4} - \sum_{i<j}\deriv{}{r_{ij}}{} \\
 & - \frac{1}{4}\sum_{i<j<k}  \bigg(\frac{\bri{ij} }{ r_{ij}} \cdot \frac{\bri{ik} }{ r_{ik}} + \frac{\bri{ji} }{ r_{ji}} \cdot \frac{\bri{jk}  }{ r_{jk}} + \frac{\bri{ki} }{ r_{ki}} \cdot  \frac{\bri{kj} }{ r_{kj}} \bigg),
 \end{aligned}
\end{equation}
which means that all scalar electron-electron repulsive interaction have been replaced by a differential operator and a three electron interaction. 
It should be stressed that even if the $\mu \rightarrow 0$ limit of $\tilde{H}[\mu]$ is very unphysical, it conserves exactly the same spectrum than the usual Hamiltonian because it originates from a similarity transformation.

\subsubsection{A special case of $\tilde{H}[\mu]$ which mimics the FROGG}
\begin{figure*}
        \includegraphics[width=0.45\linewidth]{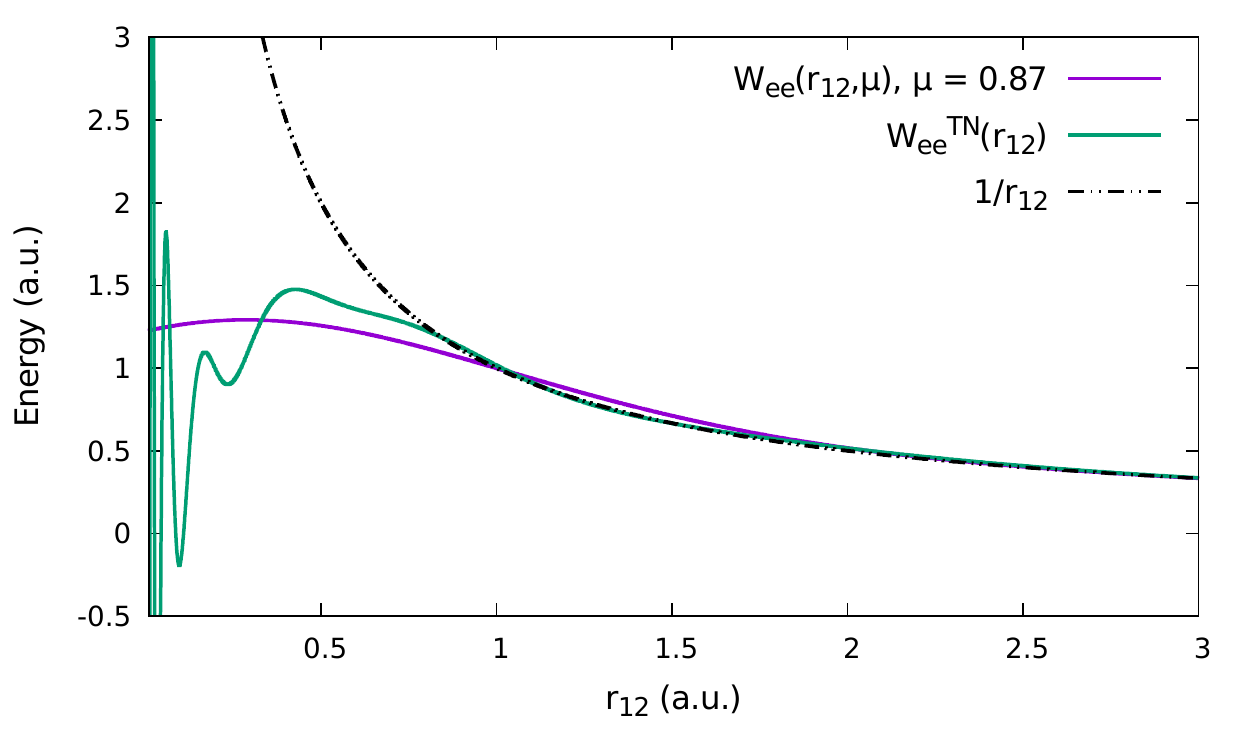}
        \caption{Shape of $\tilde{\mathcal{W}}_{ee}(r_{12},\mu)$ for $\mu=\mfrogg$ and comparison with the effective potential $\tilde{\mathcal{W}}_{ee}^\text{TN}(r_{12})$ obtained with the FROGG (see Eq. \eqref{eq:_wee_frogg}).}
 \label{fig_wee_frogg}
\end{figure*}

As the Jastrow factor $u(r_{12},\mu)$ for $\mu=\mfrogg$ is very close to the FROGG, we can also compare the scalar effective potential $\tilde{\mathcal{W}}_{ee}(r_{12},\mfrogg)$ with that obtained with the transcorrelated Hamiltonian obtained with the FROGG. 
The latter is simply defined as 
\begin{equation}
 \label{eq:tn_htilde}
 \tilde{H}_{\text{TN}} = e^{-\hat{\tau}_{\text{TN}}}H e^{\hat{\tau}_{\text{TN}}},
\end{equation}
with $\hat{\tau}_{\text{TN}} = \sum_{i<j} u^\text{TN}(r_{ij})$ and $u^\text{TN}(r_{ij})$ given in Eq. \eqref{eq_frogg}. 
This leads to the following transcorrelated Hamiltonian 
\begin{equation}
 \label{eq:tn_htilde_2}
  \tilde{H}_{\text{TN}} = H + \sum_{i<j} \bigg( \tilde{\mathcal{W}}^\text{TN}_{ee}(r_{ij}) + \tilde{t}^\text{TN} \bigg) - \sum_{i<j<k} \hat{L}^\text{TN}(\bri{1},\bri{2},\bri{3})
\end{equation}
where 
\begin{equation}
 \label{eq:_wee_frogg}
 \tilde{\mathcal{W}}_{ee}^\text{TN}(r_{12})= \frac{1}{r_{12}}-\Delta_1 u^{\text{TN}}(r_{12}) - \big(\nabla_1 u^{\text{TN}}(r_{12}) \big) ^2, 
\end{equation}
\begin{equation}
 \begin{aligned}
 \tilde{t}^\text{TN} = &\nabla_1 u^\text{TN}(r_{12}) \cdot \nabla_2 + \nabla_2 u^\text{TN}(r_{12})\cdot \nabla_1,
 \end{aligned}
\end{equation}
\begin{equation}
 \begin{aligned}
 \label{def_k}
  \hat{L}^{\text{TN}}(\bri{1},\bri{2},\bri{3})  = & \nabla_1 u^\text{TN}(r_{12}) \cdot \nabla_1 u^\text{TN}(r_{13}) \\ 
                                     + & \nabla_2 u^\text{TN}(r_{21}) \cdot \nabla_2 u^\text{TN}(r_{23})  \\
                                     + & \nabla_3 u^\text{TN}(r_{31}) \cdot \nabla_3 u^\text{TN}(r_{32}).
 \end{aligned}
\end{equation}
As apparent from Fig. \ref{fig_wee_frogg} where $\tilde{\mathcal{W}}_{ee}^\text{TN}(r_{12})$ and $\tilde{\mathcal{W}}_{ee}(r_{12},\mfrogg)$ are represented, while the latter is perfectly smooth for all $r_{12}$, the former strongly oscillates near $r_{12}=0$ between positive and negative values and is not bounded for $r_{12}=0$. 
Such behaviour of $\tilde{\mathcal{W}}_{ee}^\text{TN}(r_{12})$ near $r_{12}=0$ comes from the fact that,
as it is made of gaussian functions, $u^{\text{TN}}(r_{12})$ is cusp-less \textit{i.e.} 
\begin{equation}
 \lim_{r_{12}\rightarrow 0} \deriv{}{r_{12}}{}u^{\text{TN}}(r_{12}) = 0, 
\end{equation}
and as a necessary condition to obtain a non divergent scalar potential is (see Eqs. \eqref{eq:cusp_phi_0}  and \eqref{eq:cusp_phi_1} in Appendix \ref{sec:cusp}) 
\begin{equation}
 \label{eq:cusp_phi_0}
 \lim_{r_{12}\rightarrow 0} \deriv{}{r_{12}}{}u(r_{12}) = \frac{1}{2}, 
\end{equation}
$\tilde{\mathcal{W}}_{ee}^\text{TN}(r_{12})$ is necessary not bounded at $r_{12}=0$.  

Nevertheless, as it will be shown by numerical computations, the effective potentials $\tilde{\mathcal{W}}_{ee}^\text{TN}(r_{12})$ and $\tilde{\mathcal{W}}_{ee}(r_{12},\mfrogg)$ are almost identical \textit{in average}, as they numerically agree for $r_{12}>1$ and that $\tilde{\mathcal{W}}_{ee}^\text{TN}(r_{12})$ essentially oscillates around $\tilde{\mathcal{W}}_{ee}(r_{12},\mfrogg)$ for small values of $r_{12}$. 

\section{Detailed numerical study of $\tilde{H}[\mu]$ in the case of the helium atom}
\label{sec:total_he}
The first part of our numerical study study concerns the convergence of the eigenvalues of $\tilde{H}[\mu]$ with respect to the basis set, and more precisely the dependence on $\mu$ of such convergence. 

Considering a given basis set $\basis$ and the corresponding projector $P_\basis$, one can define the transcorrelated Hamiltonian projected on $\basis$ by
\begin{equation}
 \tilde{H}[\mu]^{\basis} = P_\basis \tilde{H}[\mu] P_\basis,
\end{equation}
and its eigenvalues $E_i^{\basis}[\mu]$ and right eigenvectors $\phiimub$ satisfy 
\begin{equation}
 \tilde{H}[\mu]^{\basis} \ket{\phimub} = E_i^{\basis}[\mu] \ket{\phiimub}. 
\end{equation}
As long as the basis set $\basis$ is incomplete, $E_i^{\basis}[\mu]$ might not coincide with $E_i$, and because of its non hermitian nature, $\tilde{H}[\mu]^{\basis}$ looses the variational principle and therefore its eigenvalues $E_i^{\basis}[\mu]$ have no reasons to be bounded from below.  

\subsection{$\mu$ dependence of the convergence of the ground state eigenvalue with respect to the basis set }
\subsubsection{Computational details}
We implemented all integrals needed for the computation of the matrix elements of $\tilde{H}[\mu]$ on a usual Gaussian orbital basis set (see Appendix \ref{sec:integrals} for more details) together with the modification of the Slater rules due to the presence of the non hermitian term in $\tilde{H}[\mu]$. We also implemented all necessary integrals to use the FROGG derived by Ten-No in Ref. \onlinecite{TenNo-CPL-00-a} in the context of the TC Hamiltonian.  
All implementation was done as a plugin of the program quantum package\cite{QP2}. 
We use the restricted Hartree-Fock (RHF) molecular orbitals to build all matrix elements of $\tilde{H}[\mu]$, and then we solve the two-body problem giving full flexibility to $\phiimub$ (\textit{i.e.} $\phiimub$ is expressed as a linear combination of all possible Slater determinants within a given basis set $\basis$) with a non hermitian eigensolver to obtain both right and left eigenvectors together with the eigenvalues. 

\subsubsection{Numerical results: ground state energies}
\label{sec:total_e}
\begin{figure*}
        \includegraphics[width=0.45\linewidth]{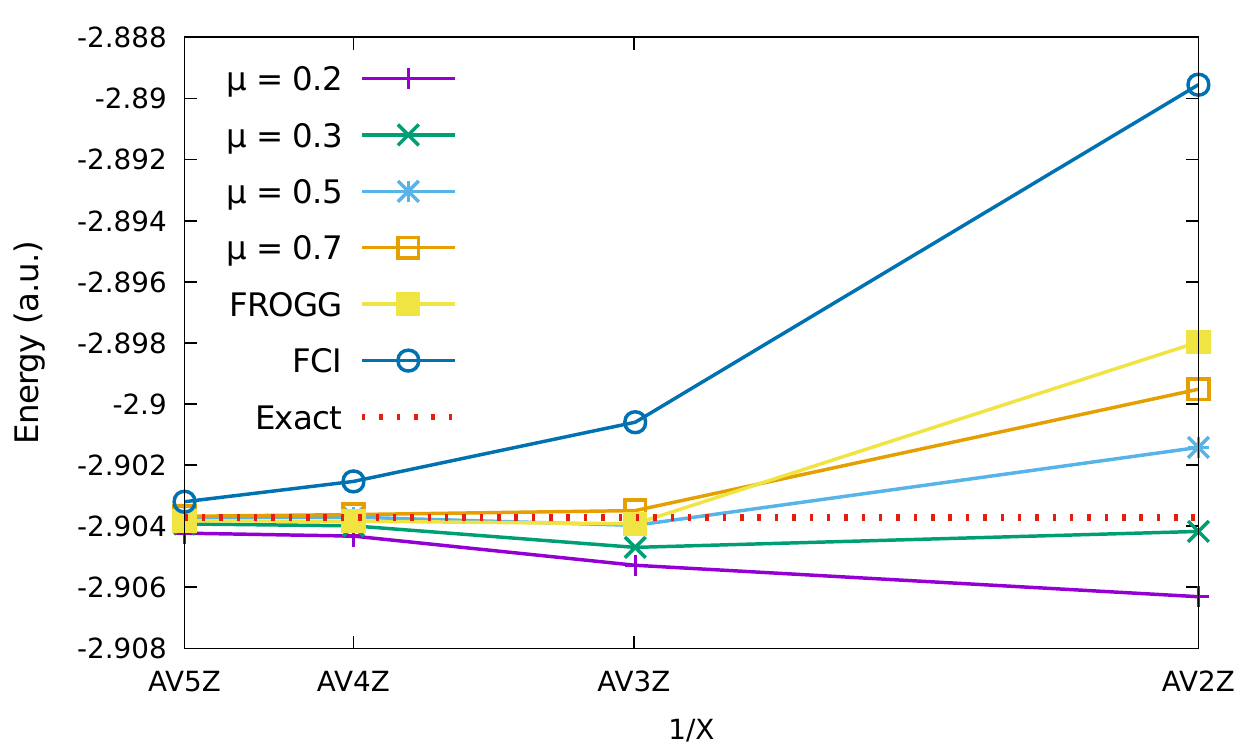}
        \includegraphics[width=0.45\linewidth]{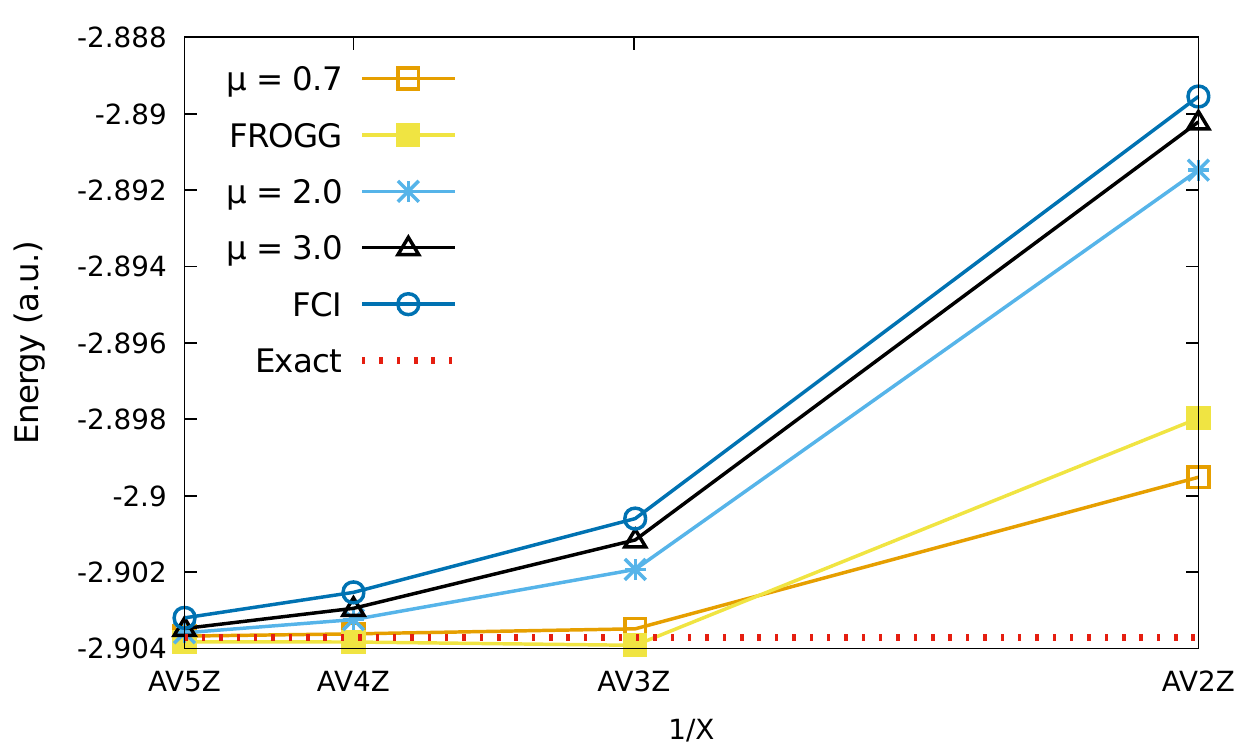}\\
        \caption{Convergence of the ground state eigenvalue of $\tilde{H}[\mu]$ with respect to the AVXZ basis set series (X=D,T,Q,5) for different values of $\mu$. 
We also compare with the results of the TC obtained with the $u(r_{12}) = u^{\text{TN}}(r_{12})$ (see Eq. \eqref{eq_frogg}), referred here after as FROGG. }
 \label{fig_conv_e_mu}
\end{figure*}

We report in Table \ref{table_conv_e_mu} the ground state eigenvalue $E_0^{\basis}[\mu]$ of $\tilde{H}[\mu]^{\basis}$ in the AVXZ basis sets (X=D,T,Q,5) for different values of $\mu$, and we report in Figure \ref{fig_conv_e_mu} the graphical representation of such data. We also report in the tables and figures the results obtained with the FROGG. 
Several observations can be done from these data. First, because of the loose of the variational principle, $E_0^{\basis}[\mu]$ can take values below the exact non relativistic energy, and the smaller the $\mu$, the more pronounced is such effect. 
Nevertheless, one can also observe that, for each $\mu$, the error with respect to the exact energy gets smaller  
when enlarging the basis set, which illustrates that in the limit of a complete basis set the eigenvalues of any TC Hamiltonian coincide with that of the physical Hamiltonian. 

Qualitatively, one can observe that there are two regimes of $\mu$: $\mu \in[0.2,0.35]$ where $E_0^{\basis}[\mu]$ is always below $E_0$ and $\mu\ge 0.7$ where $E_0^{\basis}[\mu]$ converges from above, as if the variational principle still applied. 
As $\mu$ increases, the difference between $E_0^{\basis}[\mu]$ and $E_0^{\basis}[\text{FCI}]$ in a given basis set diminishes, 
which is expected as $\lim_{\mu \rightarrow \infty} \tilde{H}[\mu] \rightarrow H$. 

Also, it is striking to see that, in a given basis set, the ground state eigenvalue of the TC Hamiltonian obtained by the FROGG and that obtained by $\tilde{H}[\mu]$ with $\mu=\mfrogg$ coincide with less than 1 mH. 
The latter results confirms that as $u(r_{12},\mfrogg) \approx u^{\text{TN}}(r_{12})$ (see Fig. \ref{fig_j_mu} and Eqs. \eqref{mu_ten_no_1} and \eqref{mu_ten_no_2}), the TC Hamiltonians obtained with the two Jastrow factor are very similar, as illustrated for instance by the effective potential in Fig. \ref{fig_wee_frogg}. 

Coming now to the speed of convergence of $E_0^{\basis}[\mu]$ with respect to the basis set, it is always faster than that of $E_0^{\basis}[\text{FCI}]$ for the whole range of values of $\mu$ selected here. 
To quantify better such observations, one can identify the basis set from which the error with respect to the exact non relativistic energy is smaller, in absolute value, than 1mH. 
For $0.3\ge\mu\ge0.4$ one can see that the error with respect to $E_0$ is never larger, in absolute value, than 1mH whatever the basis set used, which represents a strong improvement with respect to the FCI. For $0.5\ge \mu \ge 1.0$ the 1mH error threshold is reached at the aug-cc-pVTZ level and the accuracy are comparable to the FCI in the aug-cc-pV5Z basis set. For $1.6\ge \mu \ge 3.0$, such accuracy is reached at the aug-cc-pVQZ level, whenever such accuracy is reached at the aug-cc-pV5Z level using the regular FCI Hamiltonian. 

From the study of the ground state of the helium atom, one can conclude that i) the use of a simple Jastrow factor such as $u(r_{12};\mu)$ within the transcorrelated framework can strongly improve the speed of convergence of the total energy of the helium atom with a quite wide range of values of $\mu$, 
ii) the results obtained with $\mu=\mfrogg$ reproduce essentially that obtained with the FROGG, 
iii) for $\mu \in[ 0.35,0.7]$ the results obtained are always better than that obtained with the FROGG. 

\begin{table*}
\caption{Ground state eigenvalue (in Hartree) of $\tilde{H}[\mu]$ for the He atom with the AVXZ (X=D,T,Q,5) basis sets for different values of $\mu$, and error (in mH) with respect to the exact non relativistic energy. 
We also compare with the results of the TC obtained with the $u(r_{12}) = u^{\text{TN}}(r_{12})$ (see Eq. \eqref{eq_frogg} and Ref. \onlinecite{TenNo-CPL-00-a}), referred here as FROGG, and the value of $\mfrogg=0.87$ is defined in Eq. \eqref{mu_ten_no_1}. The exact non relativistic energy was obtained from Ref. \onlinecite{Dav-ions-PRA-91}. 
 }
\begin{ruledtabular}
\begin{tabular}{llllllllllll}
 Basis/$\mu$ & 0.2                   & 0.3                   & 0.35                  & 0.5                  & 0.7                 & $\mfrogg=0.87$    \\
\hline                                                                                                                                                  
 AVDZ        &    -2.906309 (-2.58)  &    -2.904172 (-0.45)  &    -2.903317 (0.41)   &    -2.901420 (2.3)   &    -2.899516 (4.21) &  -2.897927 ( 5.80)    \\
 AVTZ        &    -2.905278 (-1.55)  &    -2.904691 (-0.97)  &    -2.904468 (-0.74)  &    -2.903969 (-0.24) &    -2.903488 (0.24) &  -2.903214 ( 0.51)    \\
 AVQZ        &    -2.904325 (-0.6)   &    -2.903986 (-0.26)  &    -2.903879 (-0.15)  &    -2.903702 (0.02)  &    -2.903620 (0.1)  &  -2.903589 ( 0.13)     \\
 AV5Z        &    -2.904229 (-0.51)  &    -2.903928 (-0.2)   &    -2.903834 (-0.11)  &    -2.903702 (0.02)  &    -2.903676 (0.05) &  -2.903678 (-0.1)     \\
\hline
 Basis/$\mu$ & 1.0                   & 1.6                   & 2.0                   & 3.0                  & $\infty$ (FCI)       &     $\frogg$        \\
\hline                                                                                                                                                   
 AVDZ        &    -2.896734 (6.99)   &    -2.892773 (10.95)  &    -2.891477 (12.25)  &    -2.890212 (13.51) &    -2.889548 (14.18) &    -2.897972 (5.75 )\\
 AVTZ        &    -2.903069 (0.66)   &    -2.902421 (1.3)    &    -2.901932 (1.79)   &    -2.901159 (2.56)  &    -2.900598 (3.13)  &    -2.903919 (-0.20)\\
 AVQZ        &    -2.903558 (0.17)   &    -2.903359 (0.36)   &    -2.903244 (0.48)   &    -2.902951 (0.77)  &    -2.902534 (1.19)  &    -2.903833 (-0.11)\\
 AV5Z        &    -2.903675 (0.05)   &    -2.903634 (0.09)   &    -2.903592 (0.13)   &    -2.903479 (0.24)  &    -2.903201 (0.52)  &    -2.903825 (-0.10)\\
\hline                                                                                                                                                   
\multicolumn{7}{c}{Exact non relativistic: -2.903724}   \\
\end{tabular}
\end{ruledtabular}
\label{table_conv_e_mu}
\end{table*}

\subsection{Study of the ground state eigenvector in real space}
\label{sec:he_real_space}
In Sec. \ref{sec:total_e} we shown the improvement of the speed of convergence of the ground state energies $E_0^{\basis}[\mu]$ of the helium atom, which was effective for a quite large range of $\mu$ starting from $\mu=0.2$ to $\mu=1.6$. 
Nevertheless, one can wonder if this good behavior results from a kind of error cancellation or if there is indeed a deeper mathematical reason for such an improvement of speed of convergence. 
To bring insights to these observations, we investigate the behaviour in real space of the ground state eigenvectors of the TC Hamiltonian and the usual physical Hamiltonian in the case of the helium atom. 

For a two-electron system, the exact ground state eigenvector $\phimu$ of $\tilde{H}[\mu]$ is directly related to the exact ground state wave function $\psiex$ by the Jastrow factor $u(r_{12},\mu)$ through 
\begin{equation}
 \psiex(\br{}_1,\br{}_2) =  \frac{1}{\sqrt{\mathcal{N}^{\mu}}} e^{u(r_{12},\mu)} \phimu(\br{}_1,\br{}_2), 
\end{equation}
where the normalization factor $\mathcal{N}^{\mu}$ is simply 
\begin{equation}
  \mathcal{N}^{\mu} = \matelem{\phimu}{e^{2 u(r_{12},\mu)}}{\phimu}.
\end{equation}
Therefore, in a given basis set $\basis$, the ground state right eigenvector $\phimub$ of $\tilde{H}[\mu]^{\basis}$ can be used to estimate the exact ground state wave function $\psiex$ through 
\begin{equation}
 \psimub(\br{}_1,\br{}_2) = \frac{1}{\sqrt{\mathcal{N}_{\basis}^{\mu}}}e^{u(r_{12},\mu)}\phimub(\br{}_1,\br{}_2),
\end{equation}
where the normalization factor $\mathcal{N}_{\basis}^{\mu}$ is simply
\begin{equation}
  \mathcal{N}_{\basis}^{\mu} = \matelem{\phimub}{e^{2 u(r_{12},\mu)}}{\phimub}.
\end{equation}
Then, the quality of a given basis set $\basis$ to represent $\tilde{H}[\mu]$  can be also quantified from the vicinity between $\psimub(\br{}_1,\br{}_2)$ and the exact wave function $\psiex(\br{}_1,\br{}_2)$. 
To illustrate these ideas, 
we plotted a cutting of $\psiex(\br{}_1,\br{}_2)$ and  $\psimub(\br{}_1,\br{}_2) $ with different basis sets and values of $\mu$. The cutting used is the following: we set two electrons on a circle of radius $r=0.5$ a.u. from the nucleus and look at the wave functions as a function of the angle $\theta_{12}$ between the $\bri{1}$ and $\bri{2}$. 
For this study, we approximate the exact ground state wave function $\psiex(\br{}_1,\br{}_2)$ of the helium atom by the wave function developed by Umrigar \textit{et. al.}\cite{UmrGon-PRA-94} which contains explicitly the $r_{12}$ coordinate and which provides a total energy accurate up to twelve digits.  

\begin{figure*}
        \includegraphics[width=0.45\linewidth]{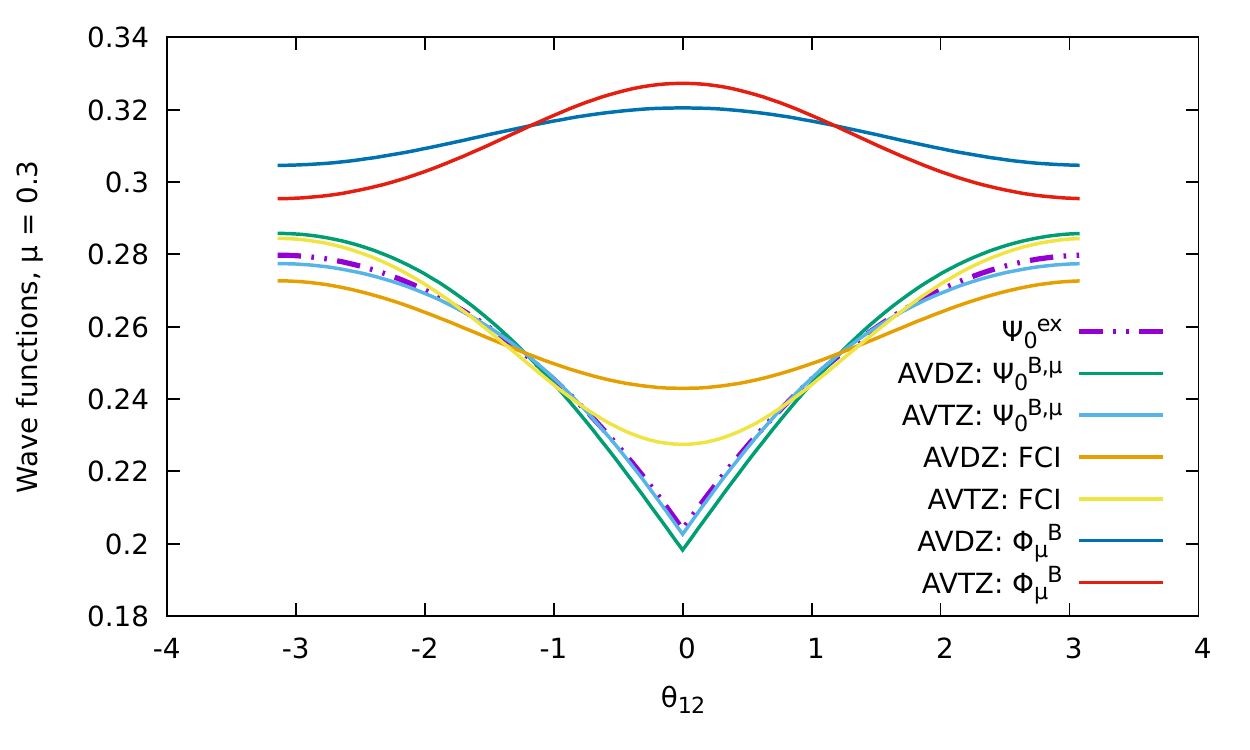}
        \includegraphics[width=0.45\linewidth]{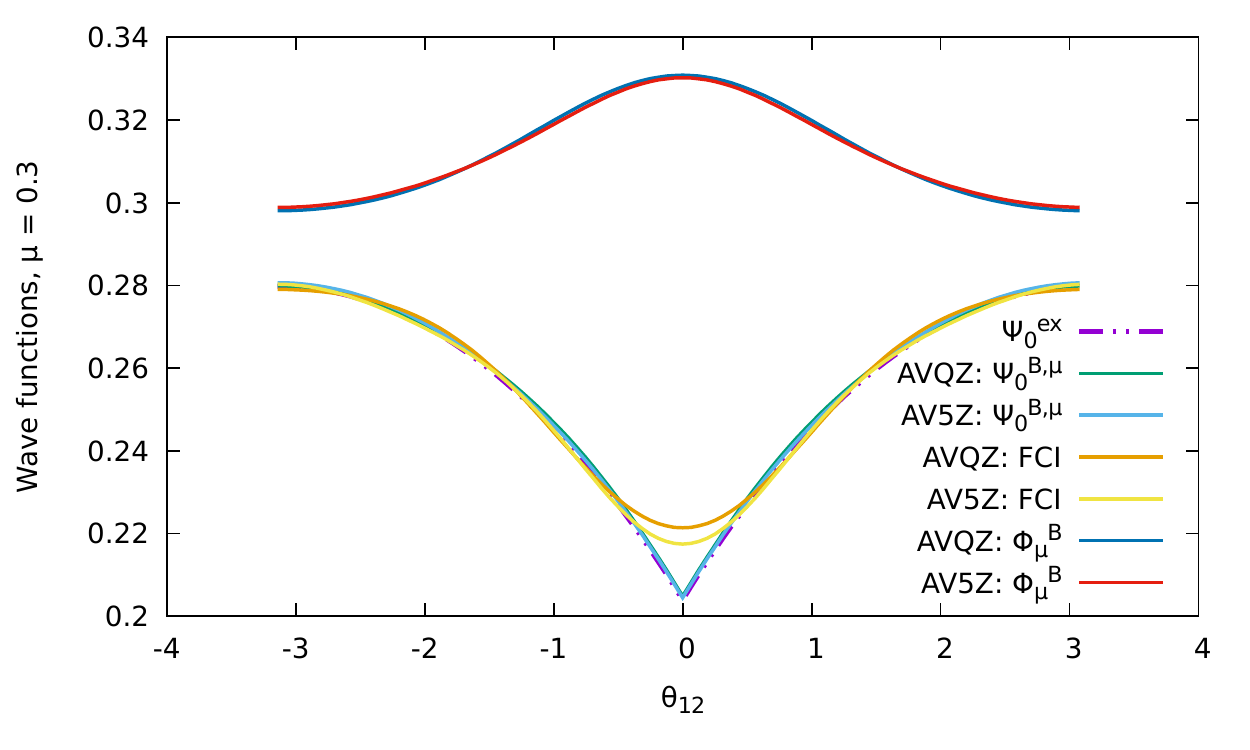}\\
        \caption{
        Helium atom, radius $r=0.5$ a.u.: approximation of the exact wave function $\psimub(\br{}_1,\br{}_2)$ and right eigenvector $\phimub(\br{}_1,\br{}_2)$ for $\mu=0.3$ in the AVDZ and AVTZ basis sets (left) and AVQZ and AV5Z (right). Comparison with the FCI wave function in the same basis sets and the estimated exact wave function $\psiex(\br{}_1,\br{}_2)$.  }
 \label{fig:mu_0.3_dz_3}
\end{figure*}

\begin{figure*}
        \includegraphics[width=0.45\linewidth]{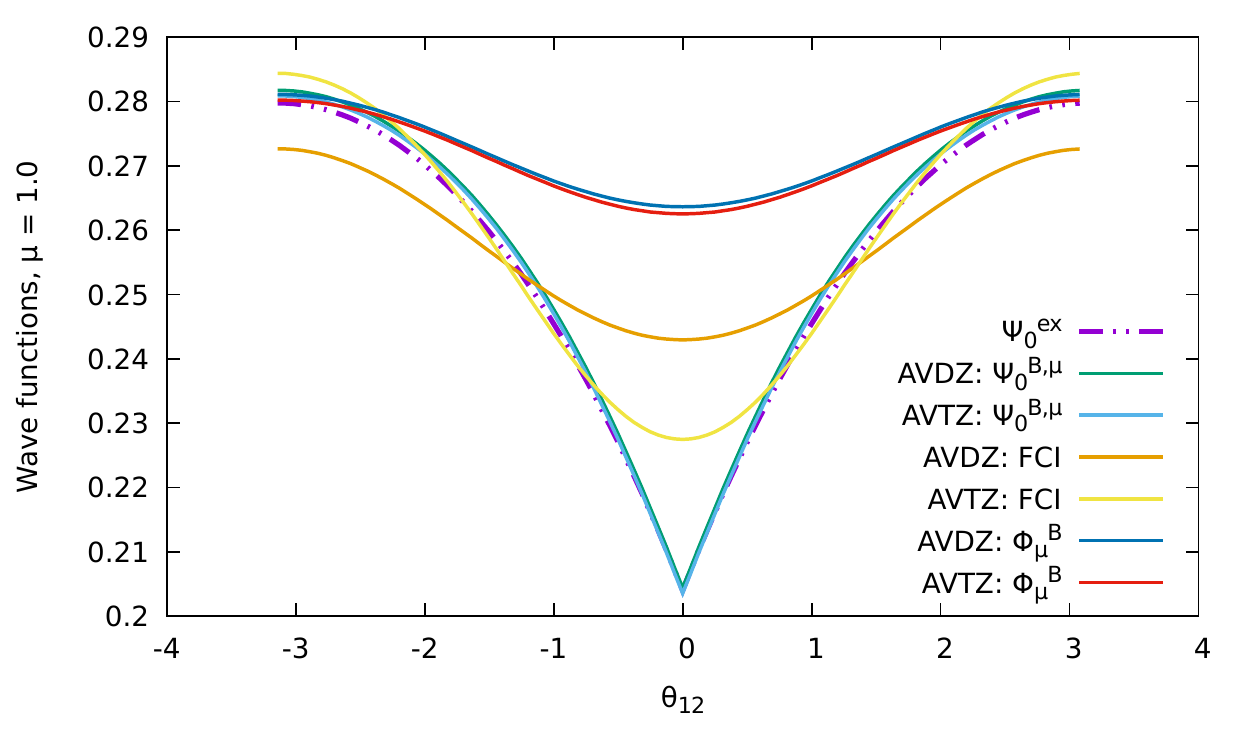}
        \includegraphics[width=0.45\linewidth]{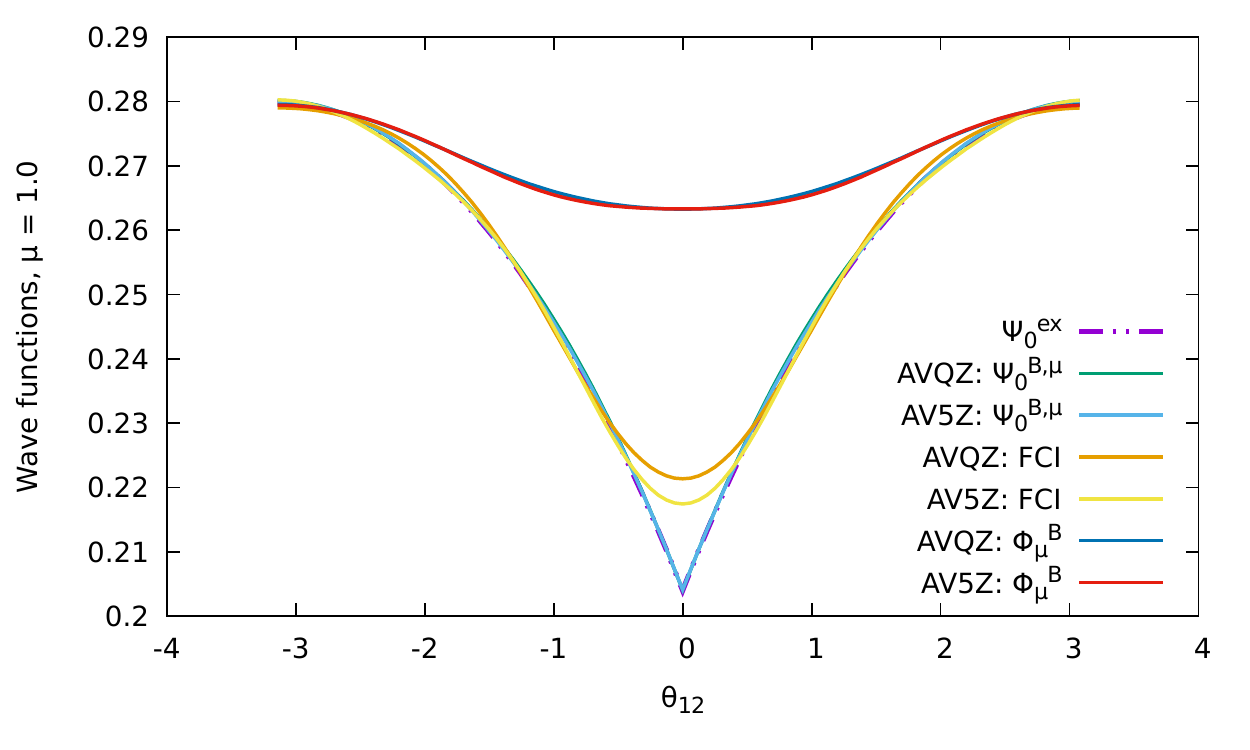}\\
        \caption{
        Helium atom, radius $r=0.5$ a.u.: approximation of the exact wave function $\psimub(\br{}_1,\br{}_2)$ and right eigenvector $\phimub(\br{}_1,\br{}_2)$ for $\mu=1.0$ in the AVDZ and AVTZ basis sets (left) and AVQZ and AV5Z (right). Comparison with the FCI wave function in the same basis sets and the estimated exact wave function $\psiex(\br{}_1,\br{}_2)$.  }
 \label{fig:mu_1.0_dz_3}
\end{figure*}

We report in Figs \ref{fig:mu_0.3_dz_3} and \ref{fig:mu_1.0_dz_3} the plot of the right eigenvector $\phimub(\br{}_1,\br{}_2)$, estimated exact wave function $\psimub(\br{}_1,\br{}_2)$ for $\mu=0.3$ and $\mu=1.0$ (respectively) and compared to the exact wave function $\psiex(\br{}_1,\br{}_2)$ and FCI wave function for $r=0.5\,a.u.$. 

From these figures, one can notice that: 
i) the right eigenvector $\phimub(\br{}_1,\br{}_2)$ converges faster than the FCI wave function with respect to the basis set, 
ii) that for $\mu=0.3$ the wave function $\phimub(\br{}_1,\br{}_2)$ presents a maximum at $r_{12}=0$, whereas for $\mu=1.0$ it has a minimum at coalescence just as the FCI wave function, 
iii) the wave function $\phimub(\br{}_1,\br{}_2)$ is always larger that the FCI one when $r_{12}\approx 0$. 
iv) that $\psimub(\br{}_1,\br{}_2)$ provides a remarkably good approximation to $\psiex(\br{}_1,\br{}_2)$ from the AVTZ basis set.

\begin{figure*}
        \includegraphics[width=0.45\linewidth]{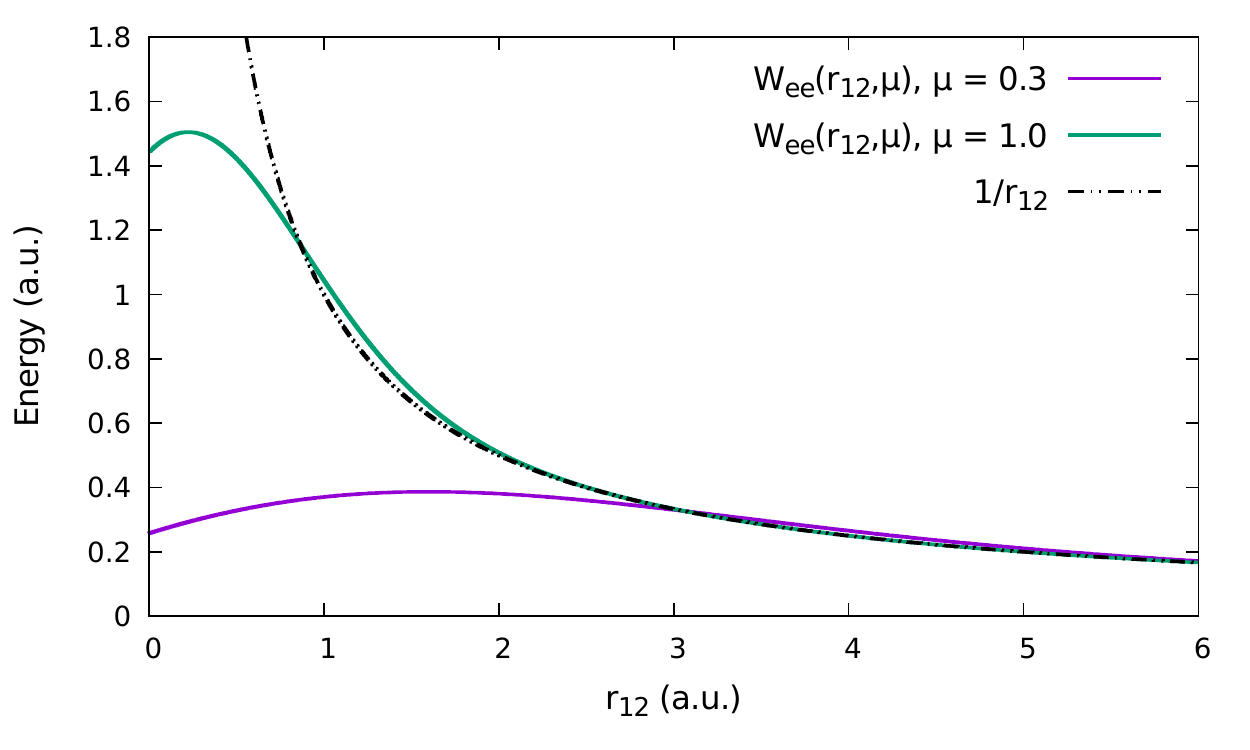}
        \caption{
        Effective potential $\tilde{\mathcal{W}}_{ee}(r_{12},\mu)$ for $\mu=0.3$ and $\mu=1.0$, compared to the usual $1/r_{12}$ Coulomb interaction.}
 \label{fig_wee_compare}
\end{figure*}

A qualitative understanding of such observations can be provided by some physical considerations regarding $\tilde{H}[\mu]$ and its dependence with $\mu$. 

i) The shape of the Jastrow factor $u(r_{12};\mu)$ leading to the transcorrelated Hamiltonian is such that as long as $\mu < \infty$, the effective potential $\tilde{\mathcal{W}}_{ee}(r_{12},\mu)$ in $\tilde{H}[\mu]$ is non divergent, and therefore its eigenfunctions do not have to fulfill the cusp condition which leads to a faster convergence with $\basis$. 
Nevertheless, one can also remark that for $\mu > 0.5$, the speed of convergence deteriorates as $\mu$ increases, even though it remains finite. This can be explained by the fact that even if $\tilde{\mathcal{W}}_{ee}(r_{12},\mu)$ remains bounded from above, when $\mu$ is large such operator is poorly represented in a finite basis set. 

ii) The fact that the wave function $\phimub(\br{}_1,\br{}_2)$ presents a maximum at $r_{12}=0$ for $\mu=0.3$ whereas it provides a minimum for $\mu=1.0$ can be explained by the shape of the effective potential $\tilde{\mathcal{W}}_{ee}(r_{12},\mu)$: the latter is much more attractive when $r_{12}\approx 0$ for $\mu=0.3$ than for $\mu = 1.0$ (see Fig. \ref{fig_wee_compare} for a graphical representation). 

iii) Similarly, as for any $\mu < \infty$ one has that $\tilde{\mathcal{W}}_{ee}(r_{12},\mu) <\frac{1}{r_{12}}$, the value of the eigenvector of $\phimub(\br{}_1,\br{}_2)$ is necessary larger than the FCI wave function at $r_{12}=0$ as in the latter the interaction is more repulsive. This implies that the on-top pair density (\textit{i.e.} the probability of finding two electrons at the same position) is necessary larger for the eigenvectors of $\tilde{H}[\mu]$ than that for the eigenvectors of $H$.

\section{Study of the helium isoelectronic series}
\label{sec:iso_elec}
Having investigated in the case of the helium atom the behaviour of both the eigenvalues and eigenvectors of $\tilde{H}[\mu]$ with both $\mu$ and the basis set used, we now study the ground state energies of the isoelectronic series of the helium atom from H$^{-}$ to Ne$^{8+}$. As seen from the previous study on the helium atom, the quality of the eigenvalues of $\tilde{H}[\mu]$ depends on the value of $\mu$ but there is a quite wide regimes of $\mu$ for which they converge significantly faster with the basis set with respect to the usual Hamiltonian. Of course, the 'optimal' range of $\mu$ might depend on the system and we investigate here different approaches to systematically find a reasonable value of $\mu$ and test it on this isoelectronic series which consist in weakly correlated systems covering a quite wide range of densities. 
As in the study of the helium atom of Sec \ref{sec:total_he}, in a given basis set $\basis$ and for a given system, we use RHF molecular orbitals and the full flexibility is given to the eigenvectors $\ket{\phimub}$. 

In order to find a systematic way to determine a reasonable value of $\mu$, one must keep in mind the physical meaning of such quantity: it has the unit of the inverse of a distance, and determines, at leading order, the depth of the hole imposed by the Jastrow factor. 
Therefore, it must have the typical scale of the inverse of correlation effects, which of course strongly depends on the system through its density for instance. In the context of RS-DFT, Toulouse \textit{et. al.}\cite{TouColSav-JCP-05} have investigated different flavour of $\mu$ varying in space through the density of the system in a given point in space. 
More specifically, they introduced (see Eq. (12) of \onlinecite{TouColSav-JCP-05}) a range separation parameter typical for correlation effects in the uniform electron gas (UEG) 
\begin{equation}
 \mursc({\bf r}) =  \frac{2\sqrt{\alpha / \pi}}{\sqrt{r_s({\bf r})}},
\end{equation}
with $\alpha = (9 \pi/4)^{-1/3}$, $r_s = (n 4 \pi /3)^{-1/3}$ and $n$ being the density. 
Such function $\mursc({\bf r})$ depends on the system and on the position in space through the density $n({\bf r})$ of the system at a given point ${\bf r}$. 
Therefore, we propose here to use the average value of $\mursc({\bf r})$ over the Hartree Fock electronic density to define the value of $\mu$ for a specific system in a given basis set:
\begin{equation}
 \label{eq:mu_av_rsc}
 \murscav = \frac{1}{N_e}\int \text{d}{\bf r} \mursc({\bf r}) n_{\text{HF}}({\bf r}) 
\end{equation}
where $n_{\text{HF}}({\bf r})$ is the HF density and $N_e$ the number of electrons in the system. 
We report in Table \ref{table_conv_e_mu_iso} the convergence with the basis set $\basis$ of the error with respect to the exact ground state energies of the isoelectronic series of the helium atom using $\murscav$ in $\tilde{H}[\mu]$. 
We also report in Table \ref{table_conv_e_mu_iso} the value of $\murscav$ in the double-zeta quality basis for each system, the value in larger basis sets varying by less than $0.1\%$. 
As one can observe from  Table \ref{table_conv_e_mu_iso}, the ground state energies of $\tilde{H}[\murscav]$ converge much faster than that of the usual Hamiltonian as the MAD for the ground state energies is of 2.64 mH, 0.48 mH and 0.26 mH in the double-, triple- and quadruple zeta quality basis sets (respectively) whereas it is of 12.16 mH, 4.71 mH and 1.64 mH for the usual Hamiltonian. Also, one can observe that, except for the low density systems such as the H${^-}$ and He species, it converges from below the exact ground state energy. Regarding now the value of $\murscav $, one can notice that it increases with the atomic number. This is expected as the density becomes more picked near the nucleus while the nuclear charge increases, which induces necessary that the typical inter electronic distance lowers with $Z$. In a similar fashion, we tried another definition of $\mu$ based on the work of Toulouse \textit{et. al.}\cite{TouColSav-JCP-05} which was proportional to $\frac{1}{r_s({\bf r})}$ and found that it leads to values of $\mu$ that are too large and therefore do not lead to significant improvement of the basis set convergence. 

Still with the aim of finding good values of $\mu$, we propose here the derivation of another approach based on the mapping of the on-top pair density of the UEG and a model built with the Jastrow factor defined by \eqref{eq:def_j}. 
Assuming a single Slater determinant ansatz for a Jastrow Slater wave function with a Jastrow factor defined in \eqref{eq:def_j} and neglecting the corresponding normalization factor, the on-top pair density is 
\begin{equation}
 n_2^\mu({\bf r}) = \frac{1}{2}\big(n({\bf r})\big)^2 e^{-\frac{1}{\sqrt{\pi}\mu}}.
\end{equation}
Also, the exact on-top pair density can be estimated from the UEG through 
\begin{equation}
 n_2^{\text{UEG}}({\bf r}) = \big(n({\bf r})\big)^2g_0(n({\bf r}))
\end{equation}
where $g_0( n)$ is the structure factor of the UEG at a given density $n$. 
Therefore, one can then find the value $\mu$ such that the two on-top pair density coincide
\begin{equation}
 \begin{aligned}
                  n_2^\mu({\bf r}) &= n_2^{\text{UEG}}({\bf r}) \\
 \Leftrightarrow  \muueg({\bf r})  &= \frac{\text{log}\bigg(2 g_0(n({\bf r}))\bigg) }{\sqrt{\pi}}.
 \end{aligned}
\end{equation}
Then, one can define an average value of $\muueg({\bf r})$ over the HF density
\begin{equation}
 \label{eq:mu_av_ueg}
 \muuegav = \frac{1}{N_e}\int \text{d}{\bf r} \muueg({\bf r}) n_{\text{HF}}({\bf r}).
\end{equation}
We report in Table \ref{table_conv_e_mu_iso} for different basis sets the error with respect to the exact ground state energy of $E_0^{\basis}[\muuegav]$ for the isoelectronic series studied here. 
From these data, one can see that the MAD is sensibly the same than that with the $\murscav$ but with a MSD of opposite sign with respect to the latter. This correlates with the fact that the values of $\muuegav$ are in general larger than that of $\murscav$, except for H$^-$. 
Because the MAD are essentially the same but that the MSD are of opposite sign, it means that there exists an optimal value of $\mu$ between $\murscav$ and $\muuegav$ which might be optimal. 
Therefore, we propose to define the average between $\murscav$ and $\muuegav$ 
\begin{equation}
 \label{eq:mu_av_ueg_rsc}
  \mursclda = \frac{\murscav   +   \muuegav }{2},
\end{equation}
and the results obtained are represented in Table \ref{table_conv_e_mu_iso}. From this data, one can clearly see that $\mursclda$ gives a better MAD and MSD as it is below 1 mH from the double-zeta quality basis set and still improve when enlarging the basis set. 

We can also compare the results obtained with $\muuegav$, $\murscav$ and $\mursclda$ to that obtained with the FROGG and $\mfrogg$, which are also reported in Table \ref{table_conv_e_mu_iso}. 
From these data one can clearly observe several things: i) while for the quite diffuse systems as H$^-$ and He, the energy obtained with the FROGG are of comparable quality that that with $\muuegav$, $\murscav$ and $\mursclda$, the situation deteriorates in a monotonic way from Li$^+$ to Ne$^{8+}$, ii) from Li$^+$ to Ne$^{8+}$, the FROGG converges from below the exact energies, and the larger the $Z$, the more pronounced is the error with respect to the exact ground state energy, specially in the double-zeta quality basis, iii) the results obtained with $\mfrogg$ show a similar behaviour with respect to that of using the FROGG, iv) the MAD obtained with the FROGG is significantly higher, specially for the double-zeta quality, than that obtained with $\muuegav$, $\murscav$ and $\mursclda$. 
These results illustrates that the Jastrow factor must adapt to the typical length-scale of the system, and therefore shows the benefit of using an easily tunable Jastrow factor such as $u(r_{12},\mu)$.

\begin{table*}
\caption{Error (in mH) with respect to the exact non relativistic energies of the ground state eigenvalue of the usual Hamiltonian and $\tilde{H}[\mu]$ for the helium isoelectronic series with Dunning basis sets basis sets for different flavour of $\mu$. For H$^{-}$ and He, the basis sets used are the aug-cc-pVXZ series (X=D,T,Q) and for the Li$^+$-Ne$^{8+}$ series, the cc-pCVXZ (X=D,T,Q) basis sets with core-valence functions were used. The mean absolute deviation (MAD) and mean signed deviation (MSD) are also reported for each basis set and method. We also report the average value of the $\mu$ considered (referred as $\muav$) in the double-zeta basis. 
The method referred as $\frogg$ is the frozen gaussian geminal introduced by Ten-No\cite{TenNo-CPL-00-a}, and the value of $\mfrogg=0.87$ is defined in Eq. \eqref{mu_ten_no_1}. The exact non relativistic energy of all systems studied here was obtained from Ref. \onlinecite{Dav-ions-PRA-91}.}

\begin{ruledtabular}
\begin{tabular}{l|rrr|r||rrr|r||rrr|r|}
                         &\multicolumn{4}{c}{H$^-$}                & \multicolumn{4}{c}{He}                  & \multicolumn{4}{c}{Li$^+$}               \\
                         &   DZ    &  TZ      &   QZ    & $\muav$&  DZ     &   TZ     &  QZ  & $\muav$   &   DZ    &   TZ     &  QZ    & $\muav$  \\
\hline 
 FCI                     &   3.72  &    1.19  &   0.61  &$+\infty$ &  14.18  &   3.13   &  1.19&$+\infty$    &  10.72  &   3.35   &  1.58  &$+\infty$  \\      
$E_0^{\basis}[\muuegav]$ &   1.69  &    0.60  &   0.38  & 0.350    &  5.28   &   0.44   &  0.12& 0.815       &  0.9    &   0.09   & -0.02  & 1.274     \\      
$E_0^{\basis}[\murscav]$ &   2.30  &    0.67  &   0.39  & 0.479    &  4.87   &   0.37   &  0.12& 0.771       &  -0.82  &  -0.38   & -0.14  & 0.980     \\      
$E_0^{\basis}[\mursclda]$&   2.03  &    0.64  &   0.39  & 0.410    &  5.07   &   0.40   &  0.12& 0.792       &  0.04   &  -0.12   & -0.08  & 1.127     \\      
$E_0^{\basis}[\frogg]$   &   3.22  &    0.91  &   0.45  &    -     &  5.75   &  -0.20   & -0.11&   -         & -4.08   &  -1.24   & -0.56  &   -       \\
$E_0^{\basis}[\mfrogg]$  &   3.12  &    0.91  &   0.46  &    -     &  5.80   &   0.51   &  0.13&   -         & -1.52   &  -0.61   & -0.21  &   -       \\
\hline              
                         &\multicolumn{4}{c}{Be$^{2+}$}            & \multicolumn{4}{c}{B$^{3+}$}            & \multicolumn{4}{c}{C$^{4+}$}    \\
                         &   DZ    &  TZ      &   QZ    & $\muav$&  DZ     &   TZ     &  QZ     & $\muav$&   DZ    &   TZ     &  QZ    & $\muav$  \\
\hline 
 FCI                     &  11.35  &   3.62   &  1.41   &$+\infty$ &  11.91  &   4.21   &  1.53   &$+\infty$ &  12.46  &   4.76   &  1.67  &$+\infty$    \\  
$E_0^{\basis}[\muuegav]$ &  1.2    &   0.17   & -0.02   &1.727     &  1.41   &   0.43   &  -0.01  &2.179     &  1.67   &   0.37   &  0.01  &2.631        \\  
$E_0^{\basis}[\murscav]$ &  -1.5   &  -0.53   & -0.17   &2.152     &  -2.16  &  -0.34   &  -0.22  &1.300     &  -2.69  &  -0.4    & -0.25  &1.434        \\  
$E_0^{\basis}[\mursclda]$&  -0.08  &  -0.15   & -0.09   &1.440     &  -0.18  &   0.09   &  -0.11  &1.740     &  -0.19  &   0.01   &  -0.1  &2.032        \\  
$E_0^{\basis}[\frogg]$   &  -6.38  &  -1.63   & -0.53   &   -      &  -8.24  &  -1.45   &  -0.49  &    -     &  -9.95  &  -1.44   &  -0.48 &     -       \\
$E_0^{\basis}[\mfrogg]$  &  -3.81  &  -1.02   & -0.32   &   -      &  -6.33  &  -1.05   &  -0.45  &    -     &  -8.48  &  -1.20   &  -0.58 &     -       \\
\hline              
                         &\multicolumn{4}{c}{N$^{5+}$}   & \multicolumn{4}{c}{O$^{6+}$}            & \multicolumn{4}{c}{F$^{7+}$}    \\
                         &   DZ    &  TZ      &   QZ     & $\muav$ &  DZ     &   TZ    &  QZ     & $\muav$&   DZ    &   TZ     &  QZ    & $\muav$  \\
\hline              
 FCI                     &  13.1   &   5.71   &  1.79    &$+\infty$  &  13.84  &   6.47  &  2.02   &$+\infty$ &  14.68  &   7.06   &  2.25   &$+\infty$   \\  
$E_0^{\basis}[\muuegav]$ &  2.06   &   0.45   &  0.01    &3.082      &  2.54   &   0.77  & -0.02   &3.533     &  3.14   &   1.15   & -0.02   &3.984       \\  
$E_0^{\basis}[\murscav]$ &  -2.97  &  -0.64   &  -0.26   &1.556      & -3.13   &  -0.70  & -0.31   &1.670     & -3.10   &  -0.54   & -0.36   &1.774       \\  
$E_0^{\basis}[\mursclda]$&  -0.01  &  -0.15   &  -0.09   &2.318      &  0.27   &  -0.04  & -0.12   &2.600     &  0.69   &   0.23   & -0.14   &2.879       \\  
$E_0^{\basis}[\frogg]$   &  -11.35 &  -1.74   &  -0.39   &    -      & -12.60  &  -1.81  & -0.34   &   -      & -13.45  &  -1.55   & -0.32   &    - \\
$E_0^{\basis}[\mfrogg]$  &   -9.98 &  -1.54   &  -0.68   &    -      & -11.09  &  -1.73  & -0.80   &   -      & -11.69  &  -1.63   & -0.90   &    - \\
\hline              
                         &\multicolumn{4}{c}{Ne$^{8+}$}  & \multicolumn{4}{c}{MAD} & \multicolumn{4}{c}{MSD}    \\
                         &   DZ    &  TZ      &   QZ     & $\muav$ &  DZ     &   TZ     &  QZ     &          &   DZ    &   TZ     &  QZ    &         \\
\hline              
 FCI                     &  15.66  &   7.61   &  2.36    &$+\infty$  & 12.16   &    4.71  &    1.64 &          &  12.16   &   4.71   &    1.64&      \\   
$E_0^{\basis}[\muuegav]$ &  3.89   &   1.57   &  0.03    &4.434      & 2.38    &    0.60  &    0.07 &          &   2.38   &   0.60   &    0.07&      \\   
$E_0^{\basis}[\murscav]$ &  -2.85  &  -0.28   & -0.35    &1.874      & 2.64    &    0.48  &    0.26 &          &  -1.21   &  -0.28   &   -0.16&      \\   
$E_0^{\basis}[\mursclda]$&  1.26   &   0.57   & -0.1     &3.1543     & 0.98    &    0.24  &    0.13 &          &   0.89   &   0.15   &   -0.03&      \\   
$E_0^{\basis}[\frogg]$   &  -13.79 &  -1.17,  & -0.26    &    -      & 8.88    &    1.31  &    0.39 &   -      & -7.09    &  -1.13   &   -0.30&    - \\
$E_0^{\basis}[\mfrogg]$  &  -11.79 &  -1.41,  & -0.90    &    -      & 7.36    &    1.16  &    0.54 &   -      & -5.58    &  -0.88   &   -0.42&    - \\
\end{tabular}
\end{ruledtabular}
\label{table_conv_e_mu_iso}
\end{table*}

\section{The ground state potential energy curve H$_2$}
\begin{figure*}
        \includegraphics[width=0.45\linewidth]{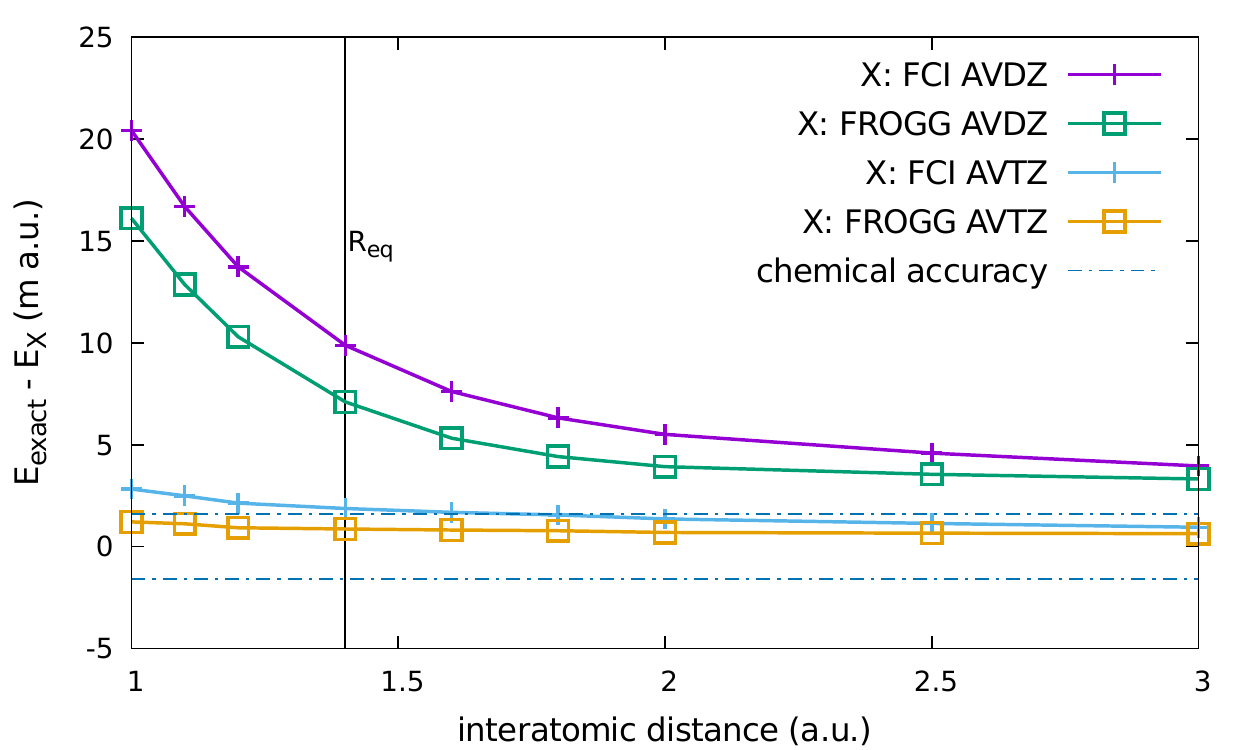}
        \includegraphics[width=0.45\linewidth]{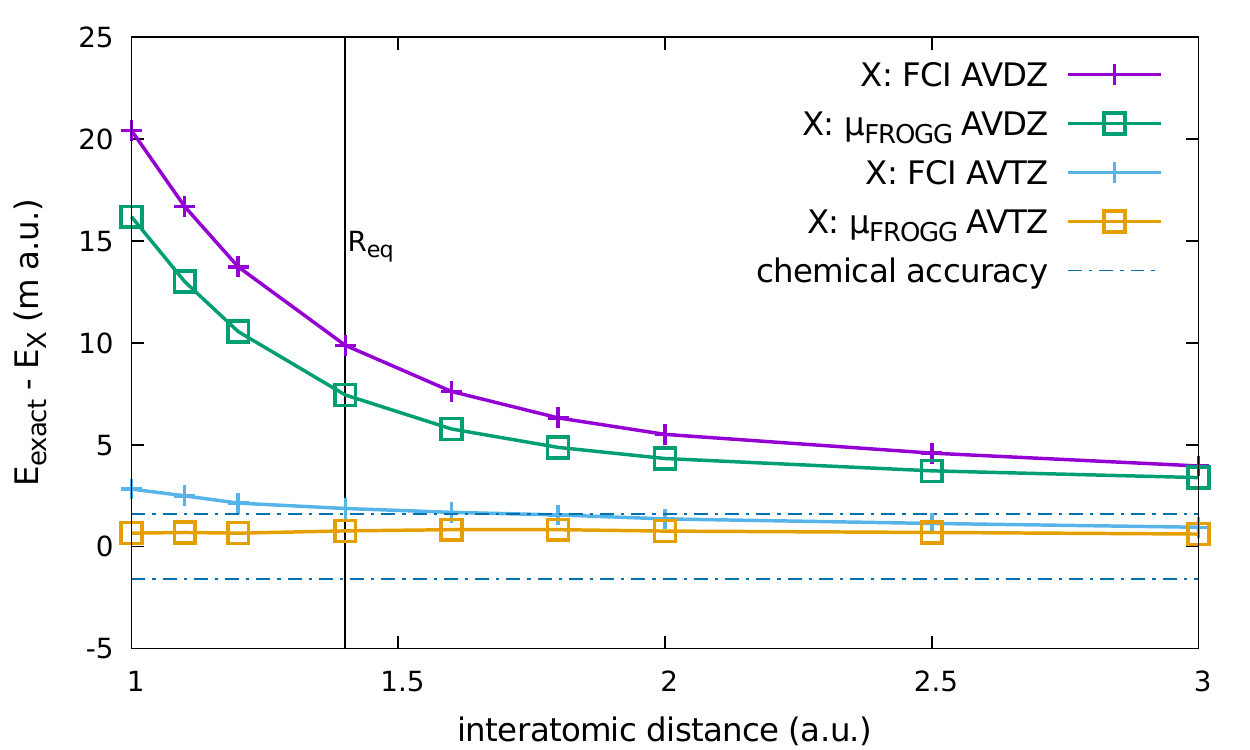}\\
        \includegraphics[width=0.45\linewidth]{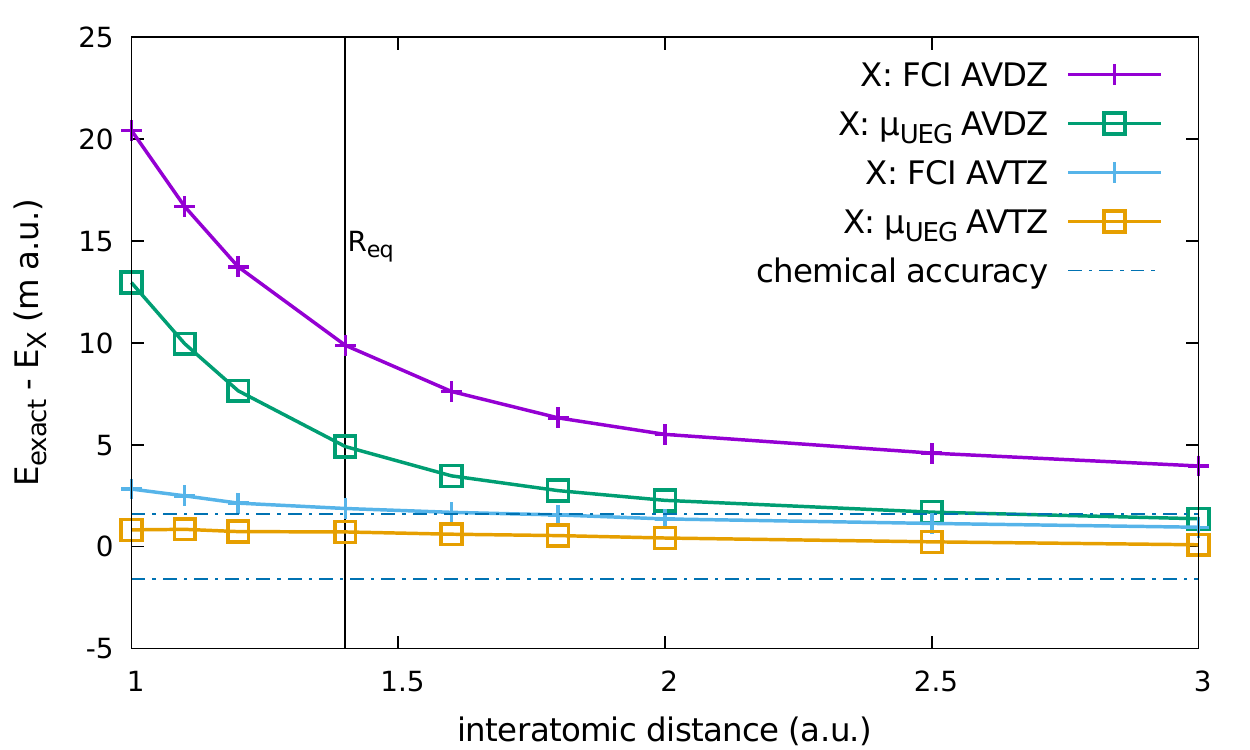}
        \includegraphics[width=0.45\linewidth]{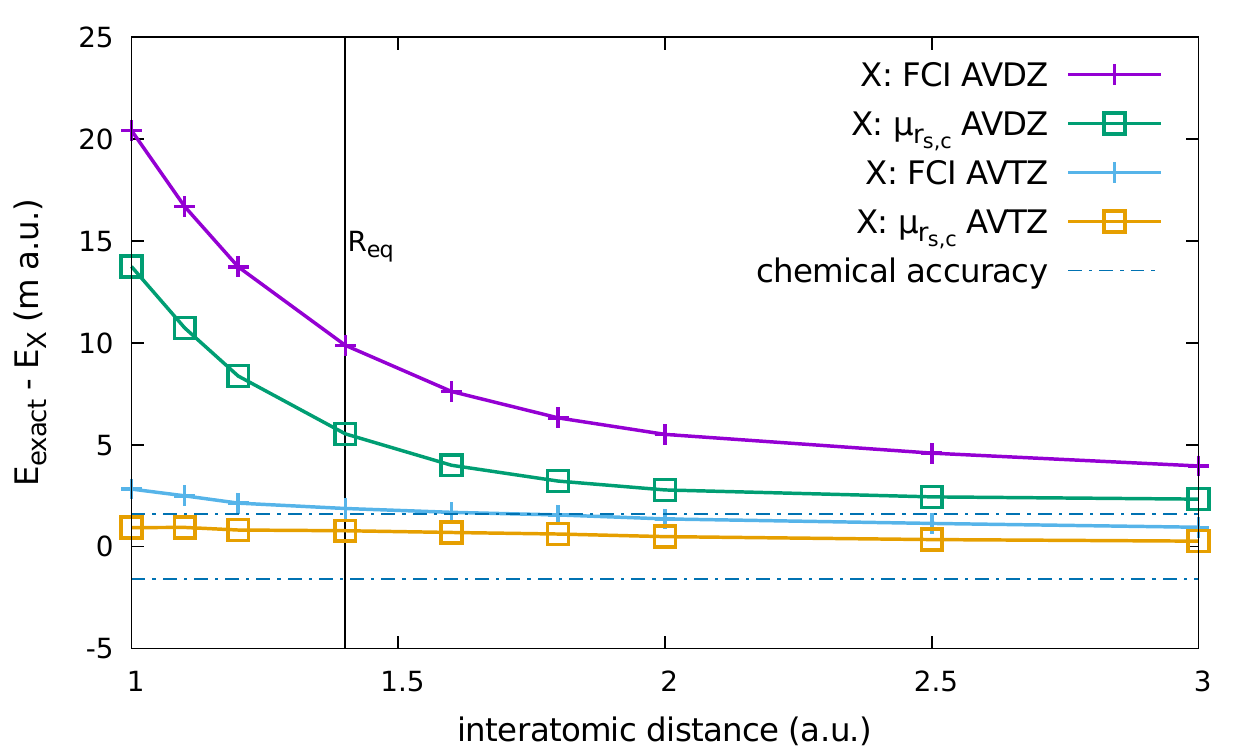}\\
        \includegraphics[width=0.45\linewidth]{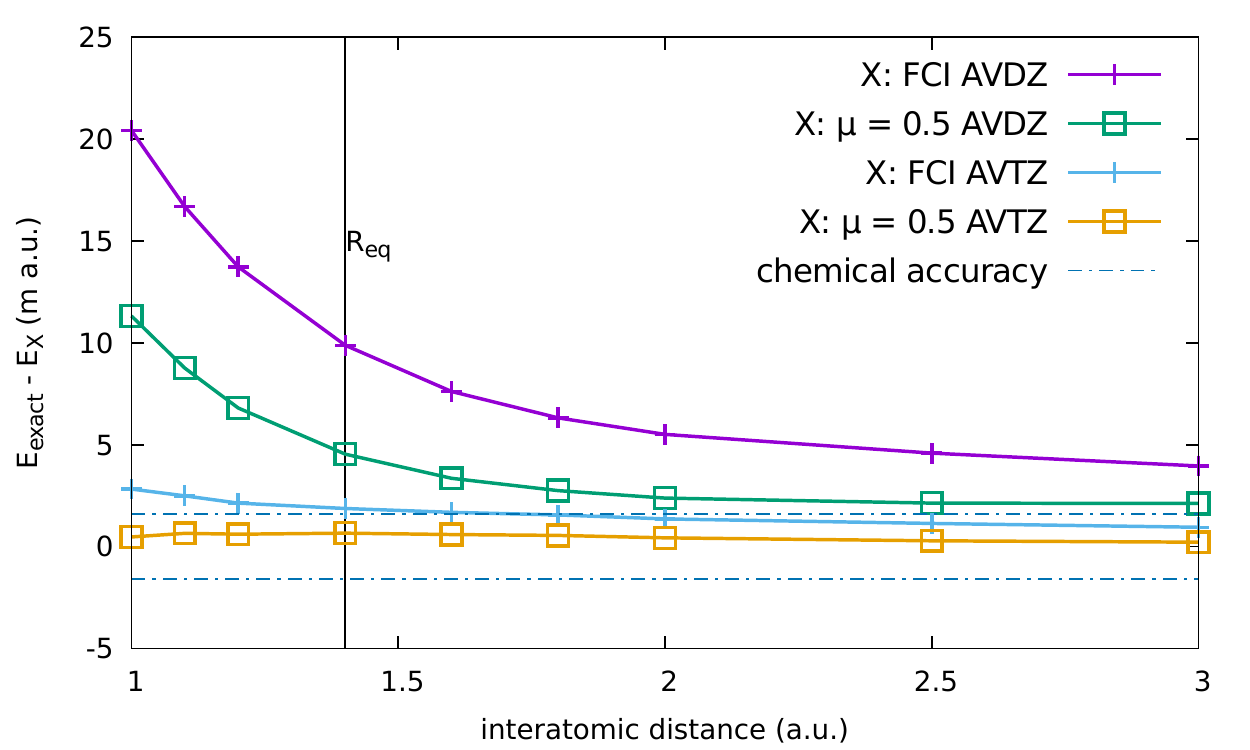}
        \includegraphics[width=0.45\linewidth]{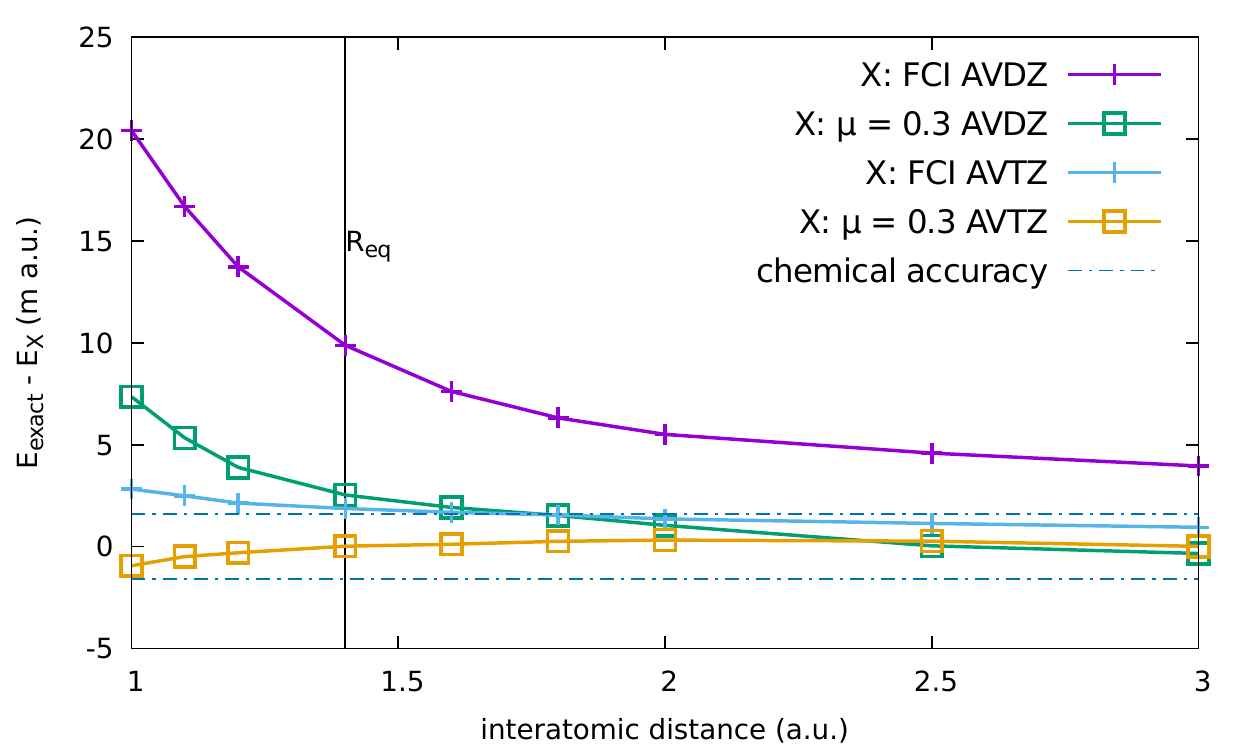}
        \caption{
        H$_2$ molecule: Error (in m a.u.) with respect to the estimated exact ground state energy for the aug-cc-pVDZ and aug-cc-pVTZ basis sets (AVDZ, AVTZ, respectively) as a function of the inter nuclear distance for different values of $\mu$, and comparison with FCI. R$_{eq}$ refers to the equilibrium distance of 1.401 a.u., and the estimated exact non relativistic potential energy curve was taken from Ref. \onlinecite{LieCle-JCP-74a}.  
}
 \label{fig:H2}
\end{figure*}

Having established in Sec \ref{sec:ht_general} the analytical form of $ \tilde{H}[\mu]$ for a general molecular system, we apply here the new TC Hamiltonian $ \tilde{H}[\mu]$ on the study of the ground state potential curve of H$_2$. 
As in the study of the helium atom of Sec \ref{sec:total_he}, in a given basis set $\basis$ and for a given geometry, we use RHF molecular orbitals and the full flexibility is given to the eigenvectors $\ket{\phimub}$. 
We report in Figs \ref{fig:H2} the difference with respect to the estimated exact ground state potential energy curve using several levels of calculations for the aug-cc-pVDZ and aug-cc-pVTZ basis sets. We do not report the values in the aug-cc-pVQZ because the FCI values are already near the CBS. 

From the Figs \ref{fig:H2}, one can observe that i) all methods provide a near chemical accuracy from the AVTZ basis set, ii) the TC Hamiltonian obtained with the FROGG and that with $\mfrogg$ provide very similar results, iii) the values obtained with $\muuegav$ and $\murscav$ provide very similar results and improve accuracy over the FROGG for the AVDZ basis set, iv) a value of $\mu = 0.5$ provides a good description in the AVDZ basis set while remaining above the exact energy in the AVTZ basis set, which is not the case for $\mu=0.3$. It should also be noticed that when the system becomes strongly correlated (\textit{i.e.} when the bond is stretched), the TC eigenvalues, whatever the Jastrow chosen, do not show any pathological behaviour. 

\section{Conclusion}
In the present paper, we derived a new form of Jastrow factor $u(r_{12},\mu)$ for the TC framework and performed the first numerical tests on a set of two electron atomic and molecular systems. 
In contrast to the FROGG introduced by Ten-No\cite{TenNo-CPL-00-a} which was obtained as a linear combination of Gaussians from a least-square fit with a fixed weighting function, the analytical form of the new Jastrow factor proposed here has been derived from the analysis of the leading order terms in $\frac{1}{r_{12}}$ of the general TC Hamiltonian, such that it mimics the long-range interaction $\text{erf}(\mu r_{12})/r_{12}$ used in the RS-DFT framework (see Sec. \ref{sec:new_j}). 
The Jastrow factor $u(r_{12},\mu)$ resulting from such mathematical conditions naturally imposes the cusp and digs the Coulomb hole over a typical length scale tuned by a unique parameter $\mu$: the smaller the $\mu$, the broader and deeper is the Coulomb hole dug by $u(r_{12},\mu)$, and in the $\mu \rightarrow \infty$ limit, $u(r_{12},\mu)=0$. 
The TC Hamiltonian $\tilde{H}[\mu]$ obtained from the similarity transformation of the usual Hamiltonian with $u(r_{12},\mu)$ is analytical and depends on a unique parameter $\mu$. 
While varying $\mu$, the physical content of $\tilde{H}[\mu]$ significantly changes: one can continuously change from a repulsive non divergent Hamiltonian at large $\mu$, to a partially attractive Hamiltonian at intermediate values of $\mu$, and eventually to a purely attractive Hamiltonian at small values of $\mu$ (see Sec. \ref{sec:ht_general}). 
Also, in the $\mu \rightarrow \infty$ limit one recovers the usual physical Hamiltonian, just as in RS-DFT. 
 
The numerical investigation of the ground state of the helium atom indicates that for values of $\mu$ ranging from 0.2 to 1.6 one can strongly improve the convergence of the computed energies with respect to the basis set (see Sec. \ref{sec:total_he}). An investigation of the right eigenvectors in real space for different values of $\mu$ have shown interesting behaviour: while for $\mu=1$ the right-eigenvectors provide a depletion of the probability of finding electrons near $r_{12}=0$, for $\mu=0.3$ one obtains an increase of such a probability. These different behaviours can be understood as the effective potential entering in the TC Hamiltonian becomes more and more attractive as $\mu$ lowers. 
Nevertheless, when the right eigenvectors are multiplied by the corresponding Jastrow factor, one obtains a very good approximation of the numerically exact ground state wave function (see Sec. \ref{sec:he_real_space}).  
Also, we found a value of $\mu \approx 0.87$ which provides a Jastrow factor very close to the FROGG, and which produces essentially the same eigenvalues than the FROGG, indicating that one can see the FROGG as a special case of the more flexible Jastrow factor $u(r_{12},\mu)$. 

In order to study the performance of the present approach on a broader set of two-electron systems, we investigated the ground state energy of the helium isoelectronic series from H$^-$ to Ne$^{8+}$ (see Sec. \ref{sec:iso_elec}). 
As the density considerably vary among the isoelectronic series, we propose a very simple scheme to provide a value of $\mu$ which automatically adapts to the system, the value of which increases with the nuclear charge of the atom considered. The results obtained for this system-dependent value of $\mu$ are found to be very accurate as the MAD with respect to the exact ground state energies over this isoelectronic series in double-zeta quality is about 1 mH and keeps lowering while improving the quality of the basis set. In contrast, the MAD obtained with the FROGG is much higher: it is of about 7 mH in the double-zeta quality basis set, which can be understood by the fact that the typical length scale of the correlation effect imposed by the FROGG is less adapted for core densities than for typical valence densities.  

The last part of our study is dedicated to the study of the ground state energy curve of H$_2$ with different schemes, and we show that the present scheme provides an improvement of the convergence of the energy over the whole potential energy curve. 

The perspective of this work are mainly to study more realistic systems which will includes the expensive three-body terms, and to explore the possibility of having a value of $\mu$ varying in space. 

\section{Acknowledgment}
The author would like to thank Julien Toulouse and Andreas Savin for stimulating discussions. 

\section{Data availability}
The data that supports the findings of this study are available within the article and from the corresponding author upon reasonable request. 

\section{Appendix}
\subsection{Analytical form for the operator $\hat{K}(\bri{1},\bri{2},\mu) $}
\label{k_l_appendix}
To compute the operators $\hat{K}(\bri{1},\bri{2},\mu)$ one needs to compute first the gradient of $u(r_{12},\mu)$ with respect to $\bri{1}$ which reads 
\begin{equation}
 \label{eq:nabla_u}
 \begin{aligned}
 \nabla_1 u(r_{12},\mu) = \frac{1 - \text{erf}(\mu r_{12})}{2 r_{12}} \big( \br{}_1 - \br{}_2 \big).
 \end{aligned}
\end{equation}
Then, the term $\big(\nabla_1 u(r_{12},\mu) \big) ^2 + \big(\nabla_2 u(r_{12},\mu) \big) ^2$ is simply 
\begin{equation}
 \begin{aligned}
 \big(\nabla_1 u(r_{12},\mu) \big) ^2 + \big(\nabla_2 u(r_{12},\mu) \big) ^2 = \frac{\bigg(1 - \text{erf}(\mu r_{12}) \bigg)^2}{2}.
 \end{aligned}
\end{equation}
The non hermitian operators in $\hat{K}(\bri{1},\bri{2},\mu)$ contains 
\begin{equation}
 \begin{aligned}
 \nabla_1 u(r_{12},\mu) \cdot \nabla_1  = &\frac{1 - \text{erf}(\mu r_{12})}{2 r_{12}} \\ 
                                          &\bigg( (x_1 - x_2) \deriv{}{x_1}{} + (y_1 - y_2) \deriv{}{y_1}{} + (z_1 - z_2) \deriv{}{z_1}{}\bigg),
 \end{aligned}
\end{equation}
and as $\nabla_1 u(r_{12},\mu) = - \nabla_2 u(r_{12},\mu)$,   
the total non hermitian operator in $\hat{K}(\bri{1},\bri{2},\mu)$ can be written as 
\begin{equation}
 \begin{aligned}
 \label{def_non_hermit}
& \nabla_1 u(r_{12},\mu) \cdot \nabla_1 + \nabla_2 u(r_{12},\mu) \cdot \nabla_2 = \frac{1 - \text{erf}(\mu r_{12})}{2 r_{12}} \\
& \bigg( (x_1 - x_2) \big( \deriv{}{x_1}{} - \deriv{}{x_2}{} \big) +
         (y_1 - y_2) \big( \deriv{}{y_1}{} - \deriv{}{y_2}{} \big)  \\
&  +      (z_1 - z_2) \big( \deriv{}{z_1}{} - \deriv{}{z_2}{} \big)\bigg).
 \end{aligned}
\end{equation}
One can notice that the form of Eq. \eqref{def_non_hermit} provides an explicit form to compute analytical integrals with Gaussian types functions. 
Nevertheless, one can notice that as 
\begin{equation}
 \deriv{}{r_{12}^x}{} = \frac{1}{2} \bigg( \deriv{}{x_1}{} - \deriv{}{x_2}{} \bigg),
\end{equation}
one can write 
\begin{equation}
 \begin{aligned}
 \label{eq:nabla_i_nabla0}
& \nabla_1 u(r_{12},\mu) \cdot \nabla_1 + \nabla_2 u(r_{12},\mu) \cdot \nabla_2 = \frac{1 - \text{erf}(\mu r_{12})}{r_{12}} \big( \br{}_1 - \br{}_2 \big) \cdot \nabla_{\br{}_{12}}.
 \end{aligned}
\end{equation}
Then, introducing the spherical coordinate system for $\br{}_1 - \br{}_2$ in terms of ${\bf e}_u= \frac{\br{}_1 - \br{}_2}{r_{12}}$ and the angle $\theta$ and $\phi$, one can write 
\begin{equation}
 \nabla_{\br{}_{12}} = \deriv{}{r_{12}}{} {\bf e}_u + \frac{1}{r_{12}} \deriv{}{\theta}{} {\bf e}_{\theta} + \frac{1}{r_{12} \sin(\theta)} \deriv{}{\phi}{} {\bf e}_\phi,
\end{equation}
and as $\br{}_1 - \br{}_2 = r_{12} {\bf e}_u$ one obtains
\begin{equation}
 \big( \br{}_1 - \br{}_2 \big) \cdot \nabla_{\br{}_{12}} = r_{12} \deriv{}{r_{12}}{}.
\end{equation}
Therefore 
\begin{equation}
 \begin{aligned}
 \label{eq:nabla_i_nabla1}
 \nabla_1 u(r_{12},\mu) \cdot \nabla_1 + \nabla_2 u(r_{12},\mu) \cdot \nabla_2 = \bigg( 1 - \text{erf}(\mu r_{12})\bigg) \deriv{}{r_{12}}{},
 \end{aligned}
\end{equation}
which is a more compact expression than Eq. \eqref{def_non_hermit}. 

The computation of the Laplacian reads 
\begin{equation}
 \begin{aligned}
 \label{eq:d2_x1_2}
 &\Delta_1 u(r_{12},\mu) + \Delta_2 u(r_{12},\mu)
 & = 2  \bigg( \frac{1 - \text{erf}(\mu r_{12})}{r_{12}} - \frac{\mu}{\sqrt{\pi}} e^{-\big(\mu r_{12} \big)^2}  \bigg).
 \end{aligned}
\end{equation}

\subsection{Computation of integrals involving $\hat{K}(\bri{1},\bri{2})$} 
\label{sec:integrals}
The only integrals involved of $\hat{K}(\bri{1},\bri{2})$ over Gaussian functions which are not analytical are of the types 
\begin{equation}
 \label{eq:k_ijkl}
 v_{ij}^{kl} = \int \dr{1} \dr{2} \phi_i(\bri{1}) \phi_j(\bri{2}) \big( g(r_{12},\mu) \big)^2  \phi_k(\bri{1}) \phi_l(\bri{2}),  
\end{equation}
with 
\begin{equation}
 g(x, \mu)= \text{erf}( \mu x) - 1.
\end{equation}
To make integrals analytical, we first fit the function $\text{erfc}(x)$ with a simple Slater-Gaussian function 
\begin{equation}
 \text{erfc}(x) \approx h(x,\alpha,\beta,c)
\end{equation}
with 
\begin{equation}
 h(x,\alpha,\beta,c) = e^{-\alpha x - \beta x^2}
\end{equation}
and $\alpha=1.09529$ and $\beta = 0.756023$. 
Then, by posing $y=\mu x$, one obtains 
\begin{equation}
 \label{fit_erf}
 \begin{aligned}
  g(x,\mu)  \approx & e^{-\alpha \mu x - \beta (\mu x)^2}\\ 
        =& h(x,\alpha \mu, \beta \mu^2).
 \end{aligned}
\end{equation}
Therefore, one can fit $g(x)^2$ as 
\begin{equation}
 \begin{aligned}
 g(x,\mu)^2&= \bigg( 1 - \text{erf}(\mu x) \bigg)^2\\
           &= \bigg( e^{-\alpha \mu x } e^{-\beta \mu^2 x^2}\bigg)^2 \\
           &= e^{-2\alpha  \mu x } e^{-2 \beta \mu^2 x^2} \\
           &= h(x,2 \alpha \mu, 2 \beta \mu^2).
 \end{aligned}
\end{equation}
Then we fit the Slater function as a linear combination of Gaussians 
\begin{equation}
 e^{-X} = \sum_{m=1}^{N_s} c_m e^{-\zeta_m X^2}. 
\end{equation}
In the present work, we use $N_s=20$ and the $\{c_m,\zeta_m\}$ parameters are reported in Table \ref{gauss_fit}.
\begin{table}
\label{gauss_fit}
\caption{Set of coefficients $c_m$ and exponents $\zeta_m$ for the fit of $e^{-X}$}
\begin{ruledtabular}
\begin{tabular}{ll}
 $\zeta_m$ & $c_m$ \\
\hline                 
   30573.77073         & 0.00338925525  \\
   5608.452381         & 0.00536433869  \\
   1570.956734         & 0.00818702846  \\
   541.3978511         & 0.01202047655  \\
   212.4346963         & 0.01711289568  \\
   91.31444574         & 0.02376001022  \\
   42.04087246         & 0.03229121736  \\
   20.43200443         & 0.04303646818  \\
   10.37775161         & 0.05624657578  \\
   5.468807545         & 0.07192311571  \\
   2.973735292         & 0.08949389001  \\
   1.661441902         & 0.10727599240  \\
   0.9505256082        & 0.12178961750  \\
   0.5552868397        & 0.12740141870  \\
   0.3304336002        & 0.11759168160  \\
   0.1998230323        & 0.08953504394  \\
   0.1224684076        & 0.05066721317  \\
   0.07575825322       & 0.01806363869  \\
   0.04690146243       & 0.00305632563  \\
   0.02834749861       & 0.00013317513  \\
\end{tabular}
\end{ruledtabular}
\end{table}
By posing $X=\gamma x$ one can fit any Slater function as
\begin{equation}
 e^{-\gamma x} = \sum_{m=1}^{N_s} c_m e^{-\zeta_m \gamma^2 x^2}. 
\end{equation}
Eventually, the function $g(x,\mu)^2$ is obtained as a linear combination of Gaussian
\begin{equation}
 g(x,\mu)^2 \approx \sum_{m=1}^{N_s} c_m e^{-2\mu^2\big(2 \alpha \zeta_m + \beta\big) x^2},
\end{equation}
which makes then the integrals analytical
\begin{equation}
 v_{ij}^{kl} \approx \sum_{m=1}^{N_s} c_m \int \dr{1} \dr{2} \phi_i(\bri{1}) \phi_j(\bri{2}) e^{-2\mu^2\big(2 \alpha \zeta_m + \beta\big) (r_{12})^2} \phi_k(\bri{1}) \phi_l(\bri{2}).
\end{equation}
All numerical tests performed for $\mu > 0.1$ show that this fit is highly accurate. 

\subsection{Analytical form of $\hat{L}(\bri{1},\bri{2},\bri{3},\mu)$ and computation of related integrals }
We need to compute the following integral 
\begin{equation}
 \label{eq:l_ijmkln}
 \begin{aligned}
 &  \matelem{\phi_i \phi_j \phi_m}{\hat{L}(\bri{1},\bri{2},\bri{3},\mu)}{\phi_k \phi_l \phi_n} = \\
 +& \matelem{\phi_i \phi_j \phi_m}{\hat{L}_1(\bri{1},\bri{2},\bri{3},\mu)}{\phi_k \phi_l \phi_n} \\
 +& \matelem{\phi_i \phi_j \phi_m}{\hat{L}_2(\bri{1},\bri{2},\bri{3},\mu)}{\phi_k \phi_l \phi_n} \\
 +& \matelem{\phi_i \phi_j \phi_m}{\hat{L}_3(\bri{1},\bri{2},\bri{3},\mu)}{\phi_k \phi_l \phi_n} 
 \end{aligned}
\end{equation}
where 
\begin{equation}
 \hat{L}_1(\bri{1},\bri{2},\bri{3},\mu) = w_\mu(\bri{12}) w_\mu(\bri{13}) \bri{12} \cdot \bri{13}
\end{equation}
with 
\begin{equation}
 w_\mu(\bri{12}) = \frac{1 - \text{erf}(\mu r_{12})}{2 r_{12}}.
\end{equation}
Let us compute the first term of Eq. \eqref{eq:l_ijmkln}:
\begin{equation}
 \begin{aligned}
 & \matelem{\phi_i \phi_j \phi_m}{\hat{L}_1(\bri{1},\bri{2},\bri{3},\mu)}{\phi_k \phi_l \phi_n} = \\
 & \int \dr{1} \dr{2} \dr{3} \phi_i(\bri{1}) \phi_j(\bri{2}) \phi_m(\bri{3}) w_{\mu}(r_{12}) w_{\mu}(r_{13}) \\ 
 &\qquad \qquad \qquad   \bri{12} \cdot \bri{13}  \phi_k(\bri{1}) \phi_l(\bri{2}) \phi_n(\bri{3}).
 \end{aligned}
\end{equation}
Such matrix element can be further decomposed into
\begin{equation}
 \begin{aligned}
 &  \matelem{\phi_i \phi_j \phi_m}{\hat{L}_1(\bri{1},\bri{2},\bri{3},\mu)}{\phi_k \phi_l \phi_n}  \\
 =& \matelem{\phi_i \phi_j \phi_m}{\hat{L}_1^x(\bri{1},\bri{2},\bri{3},\mu)}{\phi_k \phi_l \phi_n} \\  
+ & \matelem{\phi_i \phi_j \phi_m}{\hat{L}_1^y(\bri{1},\bri{2},\bri{3},\mu)}{\phi_k \phi_l \phi_n} \\  
+ & \matelem{\phi_i \phi_j \phi_m}{\hat{L}_1^z(\bri{1},\bri{2},\bri{3},\mu)}{\phi_k \phi_l \phi_n},  
 \end{aligned}
\end{equation}
where 
\begin{equation}
 \begin{aligned}
 \label{eq:l_1x}
& \matelem{\phi_i \phi_j \phi_m}{\hat{L}_1^x(\bri{1},\bri{2},\bri{3},\mu)}{\phi_k \phi_l \phi_n} \\  
 = & \int \dr{1} \dr{2} \dr{3} \phi_i(\bri{1}) \phi_j(\bri{2}) \phi_m(\bri{3}) w_{\mu}(r_{12}) w_{\mu}(r_{13}) \\ 
 &(x_1 - x_2) (x_1 - x_3) \phi_k(\bri{1}) \phi_l(\bri{2}) \phi_n(\bri{3}).
 \end{aligned}
\end{equation}
If we define the following function 
\begin{equation}
 W_{mn}^x({\bf r})  = \int \text{d}{\bf r'} \phi_m({\bf r}') \phi_n({\bf r}') w_{\mu}({\bf r} - {\bf r'}) (x - x'),  
\end{equation}
it can be calculated easily by
\begin{equation}
 W_{mn}^x({\bf r})  = x \, w_{mn}({\bf r}) - w_{mn}^x({\bf r})
\end{equation}
with 
\begin{equation}
 w_{mn}({\bf r}) = \int \text{d}{\bf r'} \phi_m({\bf r}') \phi_n({\bf r}') w_{\mu}({\bf r} - {\bf r'}), 
\end{equation}
and 
\begin{equation}
  w_{mn}^x({\bf r}) = \int \text{d}{\bf r'} \phi_m({\bf r}') \phi_n({\bf r}') w_{\mu}({\bf r} - {\bf r'})  x',
\end{equation}
which are analytical integrals.  
Then, one can rewrite the matrix element of Eq. \eqref{eq:l_1x} as
\begin{equation}
 \begin{aligned}
 \label{eq:l_1x_2}
& \matelem{\phi_i \phi_j \phi_m}{\hat{L}_1^x(\bri{1},\bri{2},\bri{3},\mu)}{\phi_k \phi_l \phi_n} \\  
 = & \int \text{d}{\bf r} \phi_i({\bf r})  \phi_k({\bf r}) W_{mn}^x({\bf r}) W_{jl}^x({\bf r}),
 \end{aligned}
\end{equation}
Therefore, the matrix element of the $\hat{L}_1(\bri{1},\bri{2},\bri{3},\mu)$ is simply 
\begin{equation}
 \begin{aligned}
 & \matelem{\phi_i \phi_j \phi_m}{\hat{L}_1(\bri{1},\bri{2},\bri{3},\mu)}{\phi_k \phi_l \phi_n} \\
 = & \int \text{d}{\bf r} \phi_i({\bf r})  \phi_k({\bf r}) \bigg( W_{mn}^x({\bf r}) W_{jl}^x({\bf r}) + W_{mn}^y({\bf r}) W_{jl}^y({\bf r}) + W_{mn}^z({\bf r}) W_{jl}^z({\bf r})\bigg),
 \end{aligned}
\end{equation}
which can be numerically evaluated easily. 
The total matrix element of the full operator is then 
\begin{equation}
 \label{eq:l_ijmkln_final}
 \begin{aligned}
 & \matelem{\phi_i \phi_j \phi_m}{\hat{L}(\bri{1},\bri{2},\bri{3},\mu)}{\phi_k \phi_l \phi_n} \\
 = & \int \text{d}{\bf r} \phi_i({\bf r})  \phi_k({\bf r}) \bigg( W_{mn}^x({\bf r}) W_{jl}^x({\bf r}) + W_{mn}^y({\bf r}) W_{jl}^y({\bf r}) + W_{mn}^z({\bf r}) W_{jl}^z({\bf r})\bigg) \\
 + & \int \text{d}{\bf r}\phi_u({\bf r})  \phi_l({\bf r}) \bigg( W_{mn}^x({\bf r}) W_{ik}^x({\bf r}) + W_{mn}^y({\bf r}) W_{ik}^y({\bf r}) + W_{mn}^z({\bf r}) W_{ik}^z({\bf r})\bigg) \\
 + & \int \text{d}{\bf r} \phi_m({\bf r})  \phi_n({\bf r}) \bigg( W_{jl}^x({\bf r}) W_{ik}^x({\bf r}) + W_{jl}^y({\bf r}) W_{ik}^y({\bf r}) + W_{jl}^z({\bf r}) W_{ik}^z({\bf r})\bigg), 
 \end{aligned}
\end{equation}
which can be computed on the fly through a simple numerical integration.

\subsection{$r_{12} \rightarrow 0$ limits for different forms of Hamiltonians and related cusp conditions}
\label{sec:cusp}
Having established the analytical form of the transcorrelated Hamiltonian $\tilde{H}[u]$ for a general Jastrow factor $u(r_{12})$ in the case of the helium atom in Sec. \ref{sec:he_htilde}, one can then study the behaviour of $\tilde{H}[u]$ when $r_{12}\rightarrow 0$ and compare it to two other types of Hamiltonian: the usual physical Hamiltonian and that entering the RS-DFT framework. 
The analysis of the $r_{12}\rightarrow 0$ behaviour leads to the different condition that the eigenvectors of such operators must fulfill. 
It should be noticed that, although it was established in the case of the helium atom, the leading order analysis of the $r_{12} \rightarrow 0$ limit that will be carried out in that section are valid for a general $N$ electron system. 
For the sake of simplicity of the notations we focus on the ground state of these operators but the arguments are valid for any bounded eigenvectors. 

In the case of the usual Hamiltonian, the exact ground state wave function $\psiex$ must satisfy the Schroedinger equation in real space 
\begin{equation}
 H \psiex(\br{}_1,\br{}_2) = E_0 \psiex(\br{}_1,\br{}_2)\quad \forall (\br{}_1,\br{}_2).
\end{equation}
When looking at $r_{12}\approx 0$, all terms multiplying $\frac{1}{r_{12}}$ in $H \psiex(\br{}_1,\br{}_2)$ must remain finite, which, according to Eq. \eqref{def:h_c}, translates into
\begin{equation}
 \label{eq:cusp_0}
 \lim_{r_{12}\rightarrow 0}\bigg( \frac{2}{r_{12}} \deriv{}{r_{12}}{} -\frac{1}{r_{12}}\bigg) \psiex(\bd{r_1},\bd{r_2})  = c,\quad c<\infty 
\end{equation}
or equivalently by multiplying Eq. \eqref{eq:cusp_0} by $r_{12}$
\begin{equation}
 \lim_{r_{12}\rightarrow 0}\bigg( 2 \deriv{\psiex(\bd{r_1},\bd{r_2})}{r_{12}}{} -\psiex(\bd{r_1},\bd{r_2}) \bigg) = 0, 
\end{equation}
which translates into the famous cusp condition for antiparallel spins of Kato\cite{Kat-CPAM-57}, 
\begin{equation}
 \deriv{\psiex(\bd{r_1},\bd{r_2})}{r_{12}}{}\Bigr|_{r_{12}=0} = \frac{1}{2} \psiex(r_{12}=0). 
\end{equation} 

Regarding now the similarity transformed Hamiltonians, the exact ground state eigenvector must also satisfy an eigenvalue equation in real-space,
\begin{equation}
 \label{eq:hpi0_ex}
 \tilde{H}[u] \phiex(\br{}_1,\br{}_2) = E_0 \phiex(\br{}_1,\br{}_2) \quad \forall (\br{}_1,\br{}_2).
\end{equation}
Similarly to Eq. \eqref{eq:cusp_0}, when looking at $r_{12}\approx 0$, the terms multiplying $\frac{1}{r_{12}}$ in $\tilde{H}[u]\phiex(\br{}_1,\br{}_2)$ must remain finite, 
which involves also the term $\tilde{W}[u]$ in $\tilde{H}[u]$ (see Eq. \eqref{eq:def_wt}). Therefore, imposing the finiteness of the limit $r_{12}\rightarrow 0$ of Eq. \eqref{eq:hpi0_ex} translates into
\begin{equation}
 \label{eq:cusp_phi_0}
 \lim_{r_{12}\rightarrow 0}\bigg( \frac{2}{r_{12}} \deriv{}{r_{12}}{} -\frac{1}{r_{12}}\bigg[ 1 - 2 \deriv{u(r_{12})}{r_{12}}{}\bigg]\bigg) \phiex(\bd{r_1},\bd{r_2})  = c,\quad c<\infty.
\end{equation}
Then, the term $\frac{1}{r_{12}}\bigg[ 1 - 2\deriv{u(r_{12})}{r_{12}}{}\bigg]$ in Eq. \eqref{eq:cusp_phi_0} looks like an effective electron electron interaction induced by the presence of the Jastrow factor in $\tilde{H}[u]$. 
Therefore, as long as one imposes that 
\begin{equation}
 \label{eq:cusp_phi_1}
  \deriv{u(r_{12})}{r_{12}}{}\bigg|_{r_{12}=0} = \frac{1}{2},
\end{equation}
which is nothing but the cusp condition for the Jastrow factor $u(r_{12})$, the scalar term proportional to $\frac{1}{r_{12}}$ in Eq. \eqref{eq:cusp_phi_0} vanishes at $r_{12}=0$, and then one obtains a non-divergent effective electron-electron interaction. 
Within the condition of Eq. \eqref{eq:cusp_phi_1}, a sufficient condition for $\phiex(\bd{r_1},\bd{r_2})$ to fulfill the general condition of Eq. \eqref{eq:cusp_phi_0} is
\begin{equation}
 \label{eq:cusp_phi_2}
 \deriv{\phiex(\bd{r_1},\bd{r_2})}{r_{12}}{}\Bigr|_{r_{12}=0} = 0, 
\end{equation}
which implies that as long as the Jastrow factor contains the cusp conditions, the eigenfunction of $\tilde{H}[u]$ is cuspless. 
For instance, a simple Jastrow factor of the form $u(r_{12}) = \frac{1}{2} r_{12}$ releases $\phiex(\br{}_1,\br{}_2)$ from 
the constraint of fulfilling the cusp condition, and produces a non diverging effective potential for the TC Hamiltonian. 

Another form of effective Hamiltonians leading to cuspless eigenfunctions are those obtained from RS-DFT 
where the Coulomb interaction $1/r_{12}$ is split into a non-divergent long-range interaction $\text{erf}(\mu r_{12})/r_{12}$ and a complementary short-range interaction $\text{erfc}(\mu r_{12})/r_{12}$, where $\text{erf}(x)$ and $\text{erfc}(x)$ are the error and complementary error functions, respectively. 
The parameter which controls such a splitting is the so-called range separation parameter $\mu$:  RS-DFT reduces the usual WFT when $\mu \rightarrow \infty$, and $\mu \rightarrow 0$ it gives back the usual Kohn-Sham theory.  
In practice, RS-DFT introduces a self-consistent Schroedinger-like equation which must be fulfilled by an effective wave function  $\Psi^\mu$.
Applied to a two electron system and looking in regions where $r_{12}\approx 0$, the leading terms of the self-consistent Schroedinger-like equation of RS-DFT reads 
\begin{equation}
 \label{eq:cusp_psi_mu_0}
 \lim_{r_{12}\rightarrow 0}\bigg( \frac{2}{r_{12}} \deriv{}{r_{12}}{} -\frac{\text{erf}(\mu r_{12})}{r_{12}}\bigg) \Psi^\mu(\bd{r_1},\bd{r_2})  = c,\quad c<\infty,  
\end{equation}
and as
\begin{equation}
 \lim_{r_{12} \rightarrow 0} \frac{\text{erf}(\mu r_{12})}{r_{12}} = \frac{2 \mu}{\sqrt{\pi}} , 
\end{equation}
one obtains that 
\begin{equation}
 \label{eq:cusp_psi_mu_1}
 \deriv{\Psi^\mu(\bd{r_1},\bd{r_2})}{r_{12}}{}\Bigr|_{r_{12}=0} = 0, 
\end{equation}
which means that the wave functions $\Psi^\mu$, just as $\phiex$ are cuspless, because they deal with an effective non-divergent interaction. 

Therefore, one can see that there is a similarity between the eigenvectors of RS-DFT and of the transcorrelated Hamiltonian: they are both cuspless as they originate from non divergent Hamiltonians. 

\subsection{Large $\mu$ limit of $\tilde{H}[\mu]$}
\label{sec:large_mu_lim}
As mentioned in Sec. \ref{sec:h_mu_lim}, one must recover the usual physical Hamiltonian in the large $\mu$ limit. 
Nevertheless, one can notice that 
\begin{equation}
 \label{eq:lim_mu_3}
 \lim_{\mu \rightarrow \infty} \tilde{\mathcal{W}}_{ee}(r_{12})  = \frac{1}{r_{12}} + \delta(r_{12}) 
\end{equation}
with 
\begin{equation}
 \delta(x) = \lim_{\mu \rightarrow \infty} \frac{\mu}{\sqrt{\pi}} e^{-\big(\mu x \big)^2}, 
\end{equation}
being the Dirac distribution. Threfore, one could be tempted to write that 
\begin{equation}
 \begin{aligned}
 \label{eq:lim_mu_4}
 \lim_{\mu \rightarrow \infty} \tilde{H}[\mu]& = H + \delta(r_{12})  \\
                                             & \ne H,
 \end{aligned}
\end{equation}
which seems to contradict the intuitive limit of Eq. \eqref{eq:lim_mu_1}. 

Nevertheless, as Eq. \eqref{eq:lim_mu_4} deals with operators, the equality must be considered in the sense of distributions.  
Therefore, considering a generic bounded square integrable wave function $\psi(\br{}_1,\hdots,\br{}_N)$ yielding to a integrable pair density $n_2(\br{}_1,\br{}_2)$, the equality between $\tilde{H}[\mu]$ and $H$ is, in the sense of distributions, defined by the following equality 
\begin{equation}
 \label{eq:lim_mu_5}
 \begin{aligned}
& \matelem{\psi}{H}{\psi} = \lim_{\mu \rightarrow \infty} \matelem{\psi}{\tilde{H}[\mu]}{\psi} \\
& \Leftrightarrow H = \tilde{H}[\mu],
 \end{aligned}
\end{equation}
and considering the limit for of $\tilde{H}[\mu]$ (see Eq. \eqref{eq:lim_mu_4}), it implies that 
\begin{equation}
 \label{eq:lim_mu_6}
 \int \text{d}\br{}_1 \text{d}\br{}_2 n_2(\br{}_1,\br{}_2) \delta(r_{12}) = 0.
\end{equation}
To intuitively show Eq. \eqref{eq:lim_mu_6}, one can perform a series of change of variable $(\br{}_1,\br{}_2)\rightarrow (\br{}_{12},\frac{1}{2}(\br{}_1 + \br{}_2))$ and then use the spherical coordinates for $\br{}_{12}$ 
\begin{equation}
 \int \text{d}\br{}_1 \text{d}\br{}_2 n_2(\br{}_1,\br{}_2) \delta(r_{12}) = \int \text{d}r{}_{12}  \tilde{n}_2(r_{12}) \delta(r_{12}) r_{12}^2 
\end{equation}
where $\tilde{n_2}(x)$ is the function $n_2(\br{}_1,\br{}_2)$ integrated over $(\br{}_1,\br{}_2)$ with the constraint that 
$|\br{}_1-\br{}_2| = x$. 
As the pair density $n_2(\br{}_1,\br{}_2)$ remains finite and is integrable, the function $\tilde{n}_2(x)$ 
cannot diverge faster than $\frac{1}{x}$ when $x\rightarrow 0$ and therefore one obtains that 
\begin{equation}
 \int \text{d}r{}_{12}  \tilde{n}_2(r_{12}) \delta(r_{12}) r_{12}^2 = 0,
\end{equation}
which implies Eq. \eqref{eq:lim_mu_1}.

\bibliography{srDFT_SC}

\end{document}